\documentclass[aps,onecolumn,preprintnumbers,nofootinbib,superscriptaddress,11pt]{revtex4}
\usepackage{amsmath} \usepackage{graphicx} \usepackage{amsfonts}
\usepackage{array} \usepackage{amsthm} \usepackage{bm}
\usepackage{palatino} \usepackage{mathpazo} 
\usepackage{supertabular}
\usepackage{float}
\usepackage[breaklinks]{hyperref}
\usepackage{epsfig}
\usepackage{graphicx}
\usepackage{verbatim}
\usepackage{multirow}
\usepackage{enumitem}
\usepackage{subfig}

\newcommand{\be}{\begin{equation}}
\newcommand{\ee}{\end{equation}}
\newcommand{\ba}{\begin{eqnarray}}
\newcommand{\ea}{\end{eqnarray}}
\newcommand{\bal}{\begin{align}}
\newcommand{\eal}{\end{align}}

\newcommand{\bw}{\begin{widetext}}
	\newcommand{\ew}{\end{widetext}}

\usepackage[leftcaption]{sidecap}

\numberwithin{equation}{section}

\begin{document}
\title{\bf Static spherically symmetric wormholes in generalized $f(R,\phi)$
gravity}
\author{M. Zubair}
\email{mzubairkk@gmail.com; drmzubair@ciitlahore.edu.pk}
\affiliation{Department of Mathematics, COMSATS Institute of Information
Technology Lahore, Pakistan}
\author{Farzana Kousar}
\email{farzana.kouser83@gmail.com}
\affiliation{Department of Mathematics,
COMSATS Institute of Information Technology Lahore, Pakistan}
 \author{Sebastian Bahamonde}
\email{sbahamonde@lincoln.ac.uk, sbahamondebeltran@gmail.com}
\affiliation{School of Mathematics and Physics, University of Lincoln.
	Brayford Pool, Lincoln, LN6 7TS, UK}
\affiliation{University of Cambridge, Cavendish Laboratory, JJ Thomson Avenue, Cambridge CB3 0HE, UK}
\affiliation{Department of
	Mathematics, University College London, Gower Street, London, WC1E 6BT, UK}

\date{\today}

\begin{abstract}
In this paper, we examine static spherically symmetric wormhole solutions in generalized $f(R,\phi)$ gravity. To do this, we consider three different kinds
of fluids: anisotropic, barotropic and isotropic. We explore different $f(R,\phi)$ models and inspect the energy conditions for all of those three fluids. It is found that under some models in this theory, it is possible to obtain wormhole solutions without requiring exotic matter. The discussion
about the conditions where the standard energy conditions (WEC and NEC) are valid for the fluids is discussed in details. From our results and for our cases, we conclude that for anisotropic and isotropic fluids, realistic wormhole geometries satisfying the energy conditions can be constructed.
\end{abstract}

\maketitle


\section{Introduction}\label{sec1}

From theoretical and observational reasons, it is believed that General
Relativity (GR) might be an incomplete theory. During the last few decades,
considerable efforts have been made to formulate alternative theories of
gravity. In this perspective, scalar-tensor gravity theory appeared as one of
the most popular candidates. In 1955, Jordan proposed a complete
gravitational theory based on the idea that $G$ (the gravitational constant
in GR) plays the role of a gravitational scalar field in accordance with Dirac's argument in such a way that the gravitational constant should
be time-dependent \cite{5,1*}. In 1961, Brans and Dicke presented a
scalar-tensor gravity theory as an effort to incorporate the Mach's proposal
in the Einstein-Hilbert gravitational framework, the so-called Brans-Dicke
(BD) theory \cite{6}. This theory has been generalized in different ways, for
example by introducing some arbitrary potential functions for the scalar
field \cite{2*}, or considering a field$-$dependent BD parameter \cite{3*} or
introducing inverse curvature correction in BD action \cite{4*} or
considering a non-minimal coupling between the scalar field and matter
systems (chameleonic BD gravity) \cite{5*}.

Quintessence scalar-tensor theory is one of the minimally coupled theories
whose Lagrangian is the sum of Einstein-Hilbert Lagrangian plus a
contribution from the scalar field. Non-minimal couplings (NMC) between the
scalar field and the scalar curvature $R$ introduce an additional term of the
form $f(\phi)R$ (where $f(\phi)$ is a function of the scalar field $\phi$) in
the Lagrangian of quintessence models \cite{6*}. NMC models have been
employed to discuss various cosmological issues such as: scaling attractor
solutions which can provide accelerated cosmic expansion at the present time
\cite{7*}, oscillating universe \cite{8*}, reconcile cosmic strings
production with inflation \cite{9*}, discuss the phase space analysis
\cite{10*}, model a modified Newtonian dynamics able to model flat rotation
curves in galaxies \cite{11*}, discuss some cosmological constraints on weak
gravitational lensing in such theories \cite{12*} and others. In \cite{13*},
authors discussed cosmological perturbations and reconstruction method in a generalized scalar-tensor
theory, which is an extended form of $Rf(\phi)$ gravity, with the following
Lagrangian
\begin{equation}\nonumber
\mathcal{L}=\frac{1}{\kappa^2}\left(f(R,\phi)+\omega(\phi)\,\phi_{;\alpha}\,
\phi^{;\alpha}\right)+V(\phi)\,,\label{11}
\end{equation}
where now $f(R,\phi)$ depends on both $R$ and the scalar field $\phi$,
$V(\phi)$ is the energy potential and $w(\phi)$ is the Brans-Dicke function
which in general depends on $\phi$. Here, semicolon represents covariant differentiation, so that $\phi_{;\alpha}=\nabla_{\alpha}\phi$. From this Lagrangian (sometimes called
extended quintessence), many different scalar-tensor theories can be
recovered as special cases. In this extended quintessence theory, various
aspects have been discussed, for example: search of a vacuum energy that can
be a possible explanation of the data from high-redshift type-Ia
supernovae \cite{14*}, cosmological evolution in the presence of exponential
couplings \cite{15*}, study of structure formation based on NMC models
\cite{16*}, cosmological perturbations of such models in the metric and
Palatini formalisms \cite{17*}, non-linear structure formation in
cosmological models using the method of spherical collapse \cite{18*}, among
other studies.

Wormholes are hypothetical topological objects that provide a shortcut
connecting two distant regions in a space-time or bridging two distinct
universes. The study of such geometrical objects started in 1916 by Flamm
\cite{7} and then followed by the work of Einstein and Rosen in 1935
\cite{8}. In the latter work, they found a space-time solution whose geometry
consists in two mouths and a throat known as an Einstein-Rosen bridge. Misner
and Wheeler introduced the word ``wormhole" for such objects in 1957
\cite{9}. They also showed that wormholes cannot be traversable for standard
matter due to its instability. The current interest in wormholes started
after the important works done by Morris and Thorne and Yurtsever \cite{10}.
They formally presented a metric, the so-called Morris-Thorne metric, and
give some conditions in order to have a traversable wormhole. They showed
that wormholes can be traversable provided that they are supported by exotic
matter, which involves a stress energy tensor that violates the null energy
condition (NEC). There already exists an important number of works exploring
the possible existence of wormhole geometries in different physical
situations. In the literature, some attempts have been made to reduce the
impact of exotic matter and minimize the violation of energy conditions
\cite{11}. One interesting approach is the one made by alternatives theories
of gravity. The main idea of this approach lies on assuming that the matter
which supports the wormhole does not violate the energy conditions but all
the new terms coming from the theory produces this violation
\cite{22*,23*,24*}. The procedure is the following. In all of those modified
theories, it is possible to rewrite the field equations using effective
fluids defined as the sum of the standard fluid plus a new fluid which
represents all the new terms coming from the modified theory. Then, one can
impose that the standard matter fluid  satisfies the energy conditions (NEC
and WEC) but the effective fluids do not. Hence, one can say that those new
terms coming from modified gravity are the responsible of the violation of
the standard energy needed to support a traversable wormhole. Wormhole
solutions have been constructed in various modified theories such as $f(R)$
gravity \cite{22*}, $f(T)$ gravity \cite{23*}, $f(R,\mathcal{T})$ gravity
(where $\mathcal{T}$ is the trace of the energy-momentum tensor) \cite{24*},
BD theory \cite{25*}-\cite{28*}, metric-Palatini hybrid $f(R)$ \cite{29*},
scalar-tensor teleparallel gravity \cite{30*}, in Einstein-Gauss-Bonnet
gravity \cite{31*} and in others.

In \cite{25*}, Agnese and Camera found static spherically symmetric solutions
in BD theory which can describe wormhole solutions depending on the choice of
post Newtonian parameter $\gamma>1$. BD theory could admit traversable
wormhole solutions for both positive and negative values of BD parameter
($\omega<-2$ and $\omega<\infty$). In this study, the scalar field plays the
role of the exotic matter \cite{26*,27*}. Ebrahimi and Riazi \cite{28*} used
a traceless energy momentum to find two classes of Lorentizan wormhole
solutions in BD theory. The first one was obtained in a open universe whereas
the second wormhole solution was obtained for both open and closed universes.
However, the WEC is violated for these solutions. The existence of Euclidean
wormhole solutions has also been explored in BD theory and Induced Gravity
\cite{32*}.

In this paper, we are interested to explore the existence of traversable
wormhole geometries in the extended quintessence scalar-tensor theory given
by the Lagrangian (\ref{11}). We will study the conditions where the energy
conditions are satisfied for three different types of fluids: anisotropic,
isotropic and barotropic fluids. In the case of the anisotropic fluid, we
will specify the shape function to then find the appropriate regions where
the wormhole solutions exist. In the other cases (isotropic and barotropic
fluids), the shape function will be analytically and numerically found from the
field equations. This paper is organised as follows: Sec.~\ref{sec2} is
devoted to present the field equations for the Morris-Thorne metric in our
extended quintessence scalar-tensor theory. Additionally, in this section we
will find the general energy conditions for the modified equations under this
geometry. In Sec.~\ref{sec3}, we study in detail the validity of the energy
conditions for anisotropic fluid by assuming an specific shape function. Secs.~\ref{sec44} and \ref{sec55} are devoted to
find and study analytical wormholes solutions for the isotropic and
barotropic fluid respectively. Finally, in Sec.~\ref{sec4} we summarize our
main results.

\section{Wormhole Geometries in extended $f(R,\phi)$ Gravity}\label{sec2}
In this section we will present the field equations for the extended
$f(R,\phi)$ gravity in the Morris-Thorne geometry and then study its generic
properties to find out the general energy conditions. The extended
$f(R,\phi)$ theory is constructed with the Lagrangian (\ref{11}) in such a
way that its action takes the following form \cite{19*},
\begin{equation}\label{2.1}
S=\int d^{4}x\sqrt{-g}\left[\frac{1}{\kappa^2}\left(f\left(R,\phi\right)
+\omega(\phi)\,\phi_{;\alpha}\,\phi^{;\alpha}\right)+V(\phi)\right]+S_{m}\,,
\end{equation}
where $\kappa^2=8\pi G$ and $S_m$ represents the action of the matter. By
varying the above action with respect to the metric, we find the following
field equations,
\begin{equation}\label{2.2}
f_{R}\,R_{\mu\nu}-\frac{1}{2}\left(f+\omega(\phi)\,\phi_{;\alpha}
\,\phi^{;\alpha}\right)\,g_{\mu\nu}-f_{R;\mu\nu}+g_{\mu\nu}\,\Box f_{R}
+\omega(\phi)\,\phi_{;\mu}\,\phi_{;\nu}+g_{\mu\nu}V(\phi)=\kappa^2\,T_{\mu\nu}\,,
\end{equation}
where $\Box=g^{\mu\nu}\nabla_{\mu}\nabla_{\nu}$, $f_{R}=\partial f/\partial
R$ and $T_{\mu\nu}=\delta S_{m}/\delta g^{\mu\nu}$ is the energy-momentum
tensor. Additionally, by taking variations in \eqref{2.1} with respect to the
scalar field we find the modified Klein-Gordon equation,
\begin{equation}\label{2.2*}
2\,\omega(\phi)\,\Box\phi+\omega_{\phi}(\phi)\,\phi_{;\alpha}\,\phi^{;\alpha}
-f_{\phi}+V_{\phi}(\phi)=0\,.
\end{equation}
We can rewrite the field equation \eqref{2.2} in an effective form,
\begin{equation}\label{2.4}
G_{\mu\nu}=R_{\mu\nu}-\frac{1}{2}R g_{\mu\nu}=T_{\mu\nu}^{\rm eff}\,,
\end{equation}
where $T_{\mu\nu}^{\rm eff}$ is the effective energy-momentum tensor defined as
\begin{equation}\label{2.5}
T_{\mu\nu}^{\rm eff}=\frac{1}{f_{R}}\bigg[\kappa^2 T_{\mu\nu}+\frac{1}{2}
\left(f+\omega(\phi)\phi_{;\alpha}\phi^{;\alpha}-R f_{R}\right)g_{\mu \nu}
+f_{R;\mu\nu}-g_{\mu\nu}\Box f_{R}-\omega(\phi) \phi_{;\mu}\phi_{;\nu}
-g_{\mu\nu}V(\phi)~\bigg]\,.
\end{equation}
The metric which could describes static spherically symmetric wormholes is
the Morris-Thorne metric which can be written as \cite{10}
\begin{equation}\label{2.6}
ds^2=e^{a(r)}dt^2-e^{b(r)}dr^2-r^2(d\theta^2+\sin^2\theta d\phi^2)\,,
\end{equation}
where $a(r)$ represents the redshift function which depends on  the radial
coordinate $r$ and $b(r)$ is a function which  related to the shape function
$\beta(r)$ via
\begin{eqnarray}
e^{-b(r)}
=1-\frac{\beta(r)}{r}\,.
\end{eqnarray}
The shape function must satisfy the condition that at the throat $r_0$ is
equal to $\beta(r=r_0)=r_0$ and then it must increases from $r_0$ to
$\infty$. For the existence of standard wormholes, the shape function must
also satisfy the flaring-out condition which reads
\begin{eqnarray}
\frac{\beta(r) -\beta'(r)r} {\beta(r)^2}>0\,, \ \ \ \textrm{at} \ \ r=r_0\,.\label{fla}
\end{eqnarray}
The above condition can be also written in a short way, namely
$\beta'(r=r_0)<1$. In addition, to do not change the signature of the metric,
the shape function must also satisfy the condition $1-\beta(r)/r>0$.\\
Since we are interested on studying wormhole geometries for anisotropic,
isotropic and barotropic fluids, we will first derive the equation for the
most general of those fluids, i.e., the anisotropic fluid. Then, when it is
necessary, the other particular cases (barotropic and isotropic) can be
easily recovered. For an anisotropic fluid, the energy-momentum tensor is
defined as follows
\begin{equation}\label{2.3}
T_{\mu\nu}=(\rho+p_t)V_{\mu}V_{\nu}-p_{t}g_{\mu\nu}+(p_{r}-p_{t})X_{\mu}X_{\nu}\,,
\end{equation}
where $\rho$, $p_r$ and $p_t$ are the energy density, radial pressure and
lateral pressure of the fluid respectively measured in the orthogonal
direction of  the unit space-like vector in the radial vector
$X_{\mu}=e^{-b}\delta^{\mu}_{1}$. Additionally,  $V_{\mu}=e^{-a}
\delta^{\mu}_{0}$ is the 4-velocity which satisfies the conditions
$V^{\mu}V_{\mu}=1$, $X^{\mu}X_{\mu}=-1$ and also $X^{\mu}V_{\mu}=0$. \\
If we consider the above energy-momentum tensor and the Morris-Thorne metric
(\ref{2.6}), the generalized $f(R,\phi)$ field equations given by (\ref{2.2})
become
\begin{eqnarray}\label{2.7}
\kappa^2\rho&=&-e^{-b}f_{R}''+\frac{1}{2r}e^{-b}\left(rb'+4\right)f_{R}'
+\frac{1}{4r}e^{-b}\left(2ra''+r{a'}^2-ra'b'+4a'\right)f_{R}\nonumber\\
&&+\frac{1}{2}
\omega(\phi)e^{-b}{\phi'}^2-\frac{1}{2}f+V(\phi)\,,\\\label{2.8}
\kappa^2p_r&=&\frac{1}{2r}e^{-b}\left(ra'+4\right)f_{R}'-\frac{1}{4r}e^{-b}
\left(2ra''+r{a'}^2-ra'b'-4b'\right)f_{R}\nonumber\\
&&-\frac{1}{2}e^{-b}\omega(\phi)
{\phi'}^2+\frac{1}{2}f-V(\phi)\,,\\\label{2.9}
\kappa^2 p_t&=&e^{-b}f_{R}''+\frac{1}{2r}e^{-b}\left(ra'-rb'+2\right)f_{R}'
+\frac{1}{2r^2}e^{-b}\left(rb'-ra'+2e^b-2\right)f_{R}\nonumber\\
&&-\frac{1}{2}e^{-b}
\omega(\phi){\phi'}^2+\frac{1}{2}f-V(\phi)\,,
\end{eqnarray}
where primes denote differentiation with respect to the radial coordinate
$r$. In GR, wormhole geometries are supported by exotic matter which requires
the violation of NEC and WEC. In \cite{20}, Harko et al. discussed wormholes
in modified theories and showed that these geometries can be theoretically
constructed without the presence of exotic matter. In such scenario, matter
threading a wormhole satisfies the energy conditions and the additional
geometric components coming from the modified theory are the responsible of
the violation of the energy conditions. Hence, the violation of the NEC and
WEC are described in terms of the effective energy momentum tensor, i.e.,
\begin{eqnarray}
\textrm{WEC}:\ W^{\mu}W^{\nu}T_{\mu\nu}^{\rm eff}<0\,, \quad \textrm{NEC}:
k^{\mu}k^{\nu}T_{\mu\nu}^{\rm eff}<0\,,
\end{eqnarray}
for any $W^{\mu}$ time-like vector and any $k^{\mu}$ null-like vector. By
doing that, we can then impose that the matter satisfies those conditions:
\begin{eqnarray}
\textrm{WEC}: W^{\mu}W^{\nu}T_{\mu\nu}>0\,, \quad \textrm{NEC}: k^{\mu}k^{\nu}
T_{\mu\nu}>0\,.
\end{eqnarray}
Clearly, if NEC is violated then WEC will be also violated and if WEC is
valid, it does not imply that the NEC is satisfied. In the literature, this
approach has been discussed in different contexts including $f(R)$ gravity
\cite{22*}, curvature-matter couplings \cite{21}, braneworlds \cite{22},
$f(T)$ theory \cite{23*}, or in hybrid metric-Palatini $f(R)$ \cite{29*}.\\
Applying the flaring out condition (\ref{fla}), one directly notice that NEC
needs to be violated for the effective fluid. Hence, to have traversable
wormhole geometries we have must impose the conditions $\rho^{\rm
eff}+p_{r}^{\rm eff}<0$ and $\rho^{\rm eff}+p_{t}^{\rm eff}<0$. As we
discussed above, those conditions do not imply that the standard matter
violates NEC. Thus, we can then impose $\rho+p_{r}>0$ and $\rho+p_{t}>0$ to
ensure that the matter satisfies the NEC, which gives us
\begin{eqnarray}
\rho+p_r&=&-e^{-b}f''_{R}+\frac{1}{2r}e^{-b}\Big(r(a'+b')+8\Big)f'_{R}+\frac{1}
{r}e^{-b}(a'+b')f_{R}>0\,,\label{rhopr}\\
\rho+p_{t}&=&\frac{1}{2r}e^{-b}(ra'+6)f'_{R}+\frac{e^{-b}}{4 r^2}\Big[2r^2 a''
+ra'\left(2-r b'\right)+r^2 a'^2+2 r b'+4 e^b-4\Big]f_{R}>0\,.\label{rhopt}
\end{eqnarray}
Let us clarify that WEC will be valid if the above conditions are true and
also assuming that the energy condition is always positive $\rho>0$. Thus,
for the validity of WEC, we also need to impose that the right hand side in
Eq.~(\ref{2.7}) is always positive. For the specific case where there are not
tidal forces, i.e., when $a'(r)=0$, the above conditions become
\begin{eqnarray}
\rho+p_{r}&=&-e^{-b}f''_{R}+\frac{1}{2r}e^{-b}\Big(rb'+8\Big)f'_{R}+\frac{b'}
{r}e^{-b}f_{R}>0\,,\\
\rho+p_{t}&=&\frac{3}{r}e^{-b}f'_{R}+\frac{e^{-b}}{4 r^2}\Big[ 2 r b'+4 e^b
-4\Big]f_{R}>0\,.
\end{eqnarray}
Hereafter, we will consider $f(R,\phi)$ models given in a power-law way given
by \cite{20*}
\begin{equation}\label{2.11}
f(R,\phi)=\gamma R \phi^{n},
\end{equation}
where $\gamma$ and $n$ are constants.
Using this model, the field equations become
\begin{eqnarray}\label{2.12}
\nonumber 2\kappa^2\rho&=&\frac{\gamma^2e^{-3b}}{16r^3}\omega(\phi){\phi'}^2
\phi^{2n}\Big(2ra''-ra'b'+ra'^2+4a'\Big)\Big(-2r^2a''+r^2a'b'-r^2 a'^2-4ra'+4rb'\\
&+&4e^{b}-4\Big)+\frac{n\gamma e^{-b}}{r}\Big(rb'+4\Big)\phi'\phi^{n-1}
-2n\gamma e^{-b}\phi''\phi^{n-1}-2\gamma n(n-1)e^{-b}\phi'^2\phi^{n-2}\nonumber\\
&+&2\kappa^2 V(\phi)\,,\\\label{2.13}
\nonumber 2\kappa^2 p_r&=&\frac{n\gamma e^{-b}}{r}\left(ra'+4\right)\phi'
\phi^{n-1}-\frac{\gamma e^{-b}}{2r^2}\left(-2r^2a''+r^2a'b'-r^2 a'^2-4ra'+4rb'
+4e^{b}-4\right)\phi^{n}\\
&-&e^{-b}\omega(\phi)\phi'^2-\frac{\gamma e^{-b}}{2r}\left(2ra''-ra'b'+ra'^2
-4b'\right)\phi^{n}-2\kappa^2 V(\phi) \,,\\\label{2.14}
\nonumber 2\kappa^2 p_t&=&\frac{\gamma e^{-b}}{r^2}\left(-ra'+r b'+2e^{b}
-2\right)\phi^{n}+\frac{n\gamma e^{-b}}{r}\left(ra'-rb'+2\right)\phi'\phi^{n-1}
+2n\gamma e^{-b}\phi''\phi^{n-1}\\
&-&\frac{\gamma e^{-b}}{2r^2}\left(-2r^2a''+r^2 a'b'-r^2a'^2-4ra'+4rb'+4e^{b}
-4\right)\phi^{n}+2\gamma n(n-1) e^{-b}\phi'^2\phi^{n-2}\nonumber\\
&-&e^{-b}\omega(\phi)\phi'^2-2\kappa^2 V(\phi)\,.
\end{eqnarray}
Now, if we replace (\ref{2.11}) into (\ref{2.2*}) we get
\begin{eqnarray}\label{2.14*}
\nonumber\frac{dV}{d\phi}&=&-\frac{n\gamma e^{-b}}{2r^2}\left\{-2r^2a''+r^2
a'b'-r^2a'^2-4ra'+4rb'+4e^{b}-4\right\}\phi^{n-1}+e^{-b}\frac{d\omega}{d\phi}
\phi'^2+2\omega(\phi)e^{-b}\\
&\times&\left\{\phi''+\left(\frac{a'-b'}{2}+\frac{2}{r}\right)\phi'\right\}\,.
\end{eqnarray}
Additionally, we will assume the following power-law functions for the BD function \cite{25} and the scalar field \cite{26}
\begin{eqnarray}
\omega(\phi)=\omega_0 \phi^m\,, \quad \phi(r)=\phi_0\left(\frac{d}{r}
\right)^{\sigma_1}\,,
\end{eqnarray}
where $a_0$, $d$ and $\omega_0$ are constants.

\section{Anisotropic generic fluid description}\label{sec3}

This section is devoted to study wormholes supported by an anisotropic fluid
characterized by $\rho$, $p_r$ and $p_t$  without specifying any equation of
state. Our principal aim is to check the validity of the energy conditions
(WEC and NEC) for our model. To do this, we will specify the $b(r)$ radial
function as follows \cite{23*,29,30,31,32,33,34}
\begin{equation}\label{2.15}
b(r)=-\ln\left[1-\left(\frac{r_0}{r}\right)^{\sigma_2+1}\right]\,,
\end{equation}
where $\sigma_2$ is a constant and $r_0$ is the throat of the wormhole, which
gives us that the shape function is
\begin{equation}\label{2.16}
\beta(r)=r_0\Big(\frac{r_0}{r}\Big)^{\sigma_2}\,.
\end{equation}
This kind of shape function has been used widely in the literature and
satisfies all the conditions needed to have a wormhole geometry if
$\sigma_2>-1$ (see the flaring-out condition given by (\ref{fla})).
Table~\ref{Table0} shows the values that the shape function takes for
different constants $\sigma_2$.
\begin{table}[H]
\centering \caption{Some shape functions for different values of the parameter
$\sigma_2$} \label{Table0}
\begin{tabular}{lccccc} \hline
$\sigma_2$ & $\sigma_2=1$~~~ & $\sigma_2=1/2$~~~ & $\sigma_2=1/5$~~~ &
$\sigma_2=0$~~~ & $\sigma_2=-1/2$~~~ 
\\
Shape function $\beta(r)$~~~ & ${r_0}^2/r$~~~ & $r_0\sqrt{r_0/r}$~~~ &
${r_0}^{6/5}{r}^{-1/5}$~~~ & $r_0$~~~ & $\sqrt{r_0 r}$~~~ 
\\
\hline
\end{tabular}
\end{table}
Additionally, for this section we will also assume that the redshift function
is constant ($a'(r)=0$), or in other words, we will assume zero tidal forces.
Using power-law ansatz with model (\ref{2.11}) and radial function
(\ref{2.15}) into (\ref{2.14*}) and integrating we have scalar potential of
the form
\begin{eqnarray}\label{2.16*}
\nonumber V(\phi)&=&\frac{n\gamma{r_0}^{\sigma_2+1}{\sigma_1}(\sigma_2-3)
\phi^{n+\frac{\sigma_2+3}{\sigma_1}}}{2d^{\sigma_2+3}{\phi_0}^{\frac{\sigma_2
+3}{\sigma_1}}(n\sigma_1+\sigma_2+3)}-\frac{\omega_0m{\sigma_1}^3\phi^{m+2
+\frac{1}{\sigma_1}}}{d^2{\phi_0}^{2/\sigma_1}(m\sigma_1+2\sigma_1+1)}
+\frac{\omega_0m{\sigma_1}^3{r_0}^{\sigma_2+1}\phi^{m+2+\frac{\sigma_2+2}
{\sigma_1}}}{d^{\sigma_2+3}{\phi_0}^{\frac{\sigma_2+3}{\sigma_1}}(m\sigma_1
+2\sigma_1+\sigma_2+2)}\\
&+&\frac{\omega_0{\sigma_1}^2(2\sigma_1-m\sigma_1-2)\phi^{m+2+\frac{2}
{\sigma_1}}}{d^2{\phi_0}^{2/\sigma_1}(m\sigma_1+2\sigma_1+2)}+\frac{\omega_0
{\sigma_1}^2{r_0}^{\sigma_2+1}(m\sigma_1-\sigma_2+3)\phi^{m+2+\frac{\sigma_2+3}
{\sigma_1}}}{d^{\sigma_2+3}{\phi_0}^{\frac{\sigma_2+3}{\sigma_1}}(m\sigma_1
+2\sigma_1+\sigma_2+3)}+c_0\,.
\end{eqnarray}
Here, $c_0$ is an integration constant. It should be noted that the special cases $n\sigma_1+\sigma_2+3=0, m\sigma_1+2\sigma_1+1=0,  m\sigma_1+2\sigma_1+\sigma_2+2=0, m\sigma_1+2\sigma_1+2=0$ and $m\sigma_1+2\sigma_1+\sigma_2+3=0 $ will be excluded from our analysis. In the following discussion, we will study the validity of WEC and NEC for
the standard matter (see Eqs.~(\ref{rhopr}) and (\ref{rhopt})). Let us then
study different  different cases for $\sigma_2$ to study the validity of the energy conditions. To do this, we will fix $r_0=\phi_0=d=\omega_0=1$ and
$\kappa^2=8\pi$ for simplicity. Additionally, it can be noticed from the
equations that the constant $\gamma$ which appears from the model (see
\eqref{2.11}) only will change the behaviour of the wormhole depending on its
sign. Hence, we will study mainly two cases for this parameter, namely, when
$\gamma=1$ and $\gamma=-1$. Let us also divide our study into two main
theories:  Brans-Dicke and Induced Gravity.

\subsection{Brans-Dicke theory}\label{brans}
\begin{figure}[H]
\centering
\includegraphics[width=0.5\textwidth]{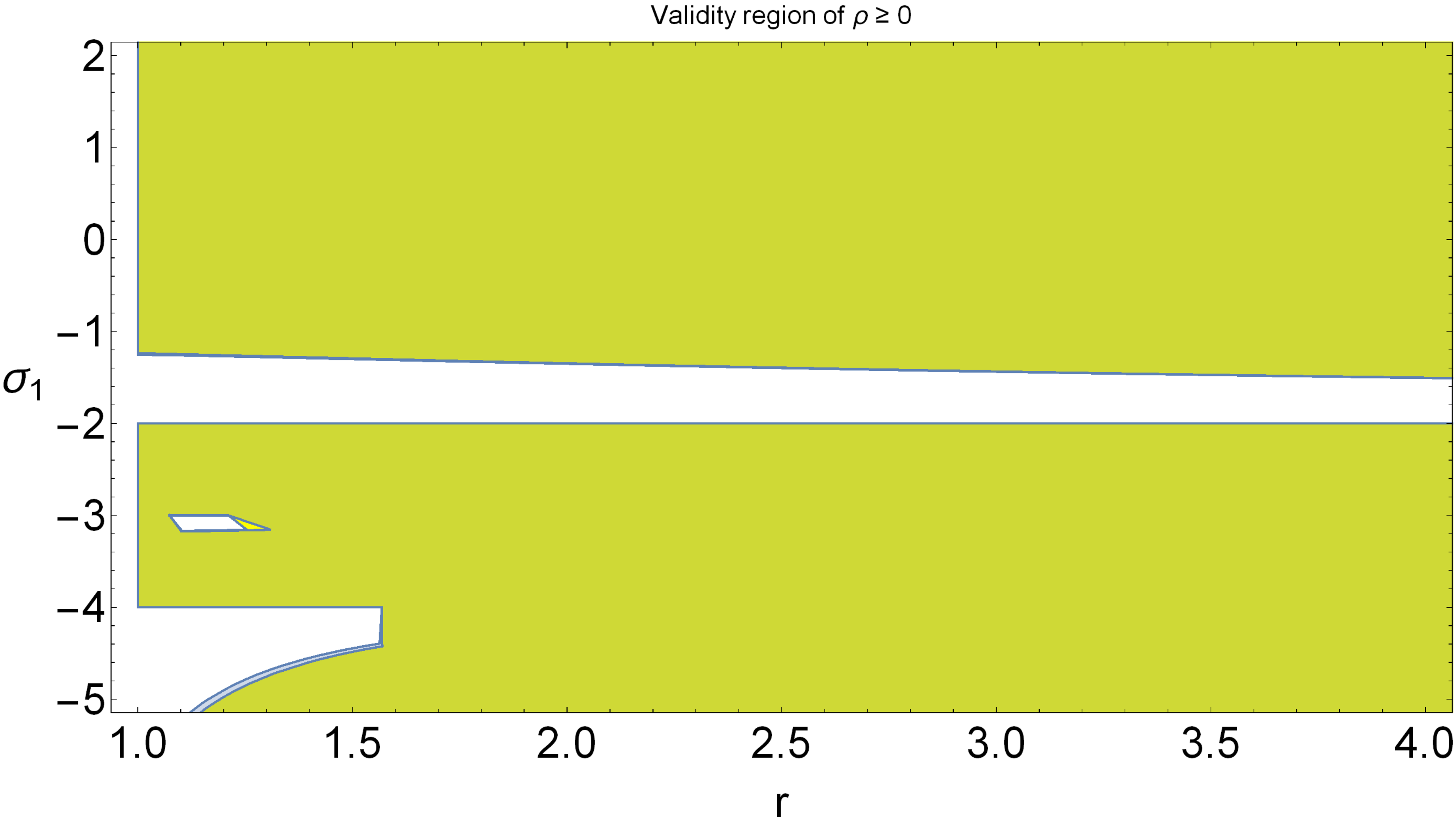}
\caption{\scriptsize{Validity of $\rho\geq0$ given by \eqref{rhorho} for the generic
anisotropic fluid in Brans-Dicke theory when $\sigma_2=1$. The yellow regions
represent the regions where $\gamma=1$ whereas the blue regions represent when
$\gamma=-1$. Therefore, the green regions represent the regions where those
two regions coincide. We have chosen the values $r_0=\phi_0=d=\omega_0=1$ and
$\sigma_2=1$.}}\label{fig:1}
\end{figure}

\begin{figure}[H]
\captionsetup{justification=raggedright}
\subfloat[NEC-1]{\includegraphics[width=0.5\textwidth]{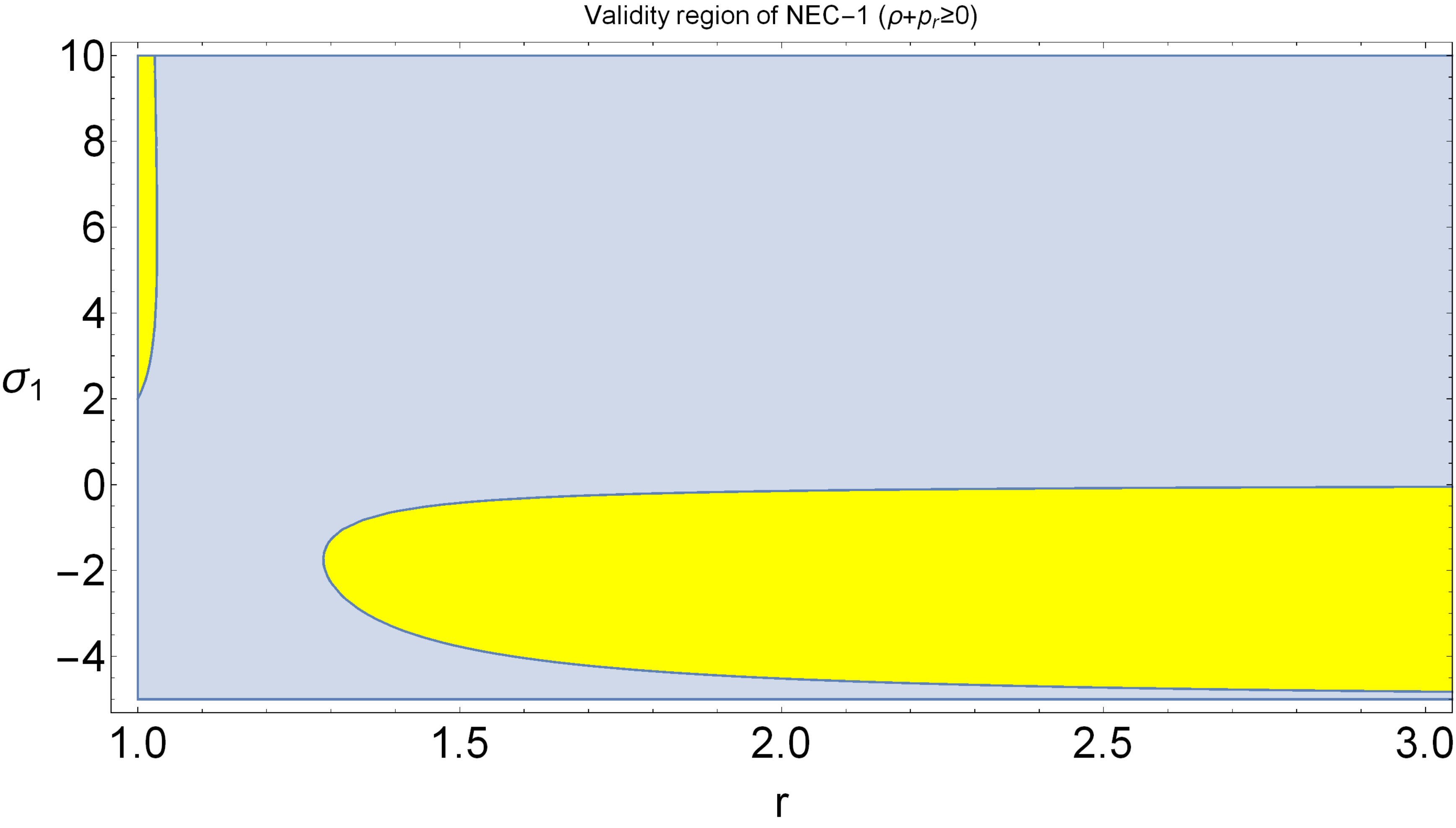}\label{fig:1a}}
\hfill
\subfloat[NEC-2]{\includegraphics[width=0.5\textwidth]{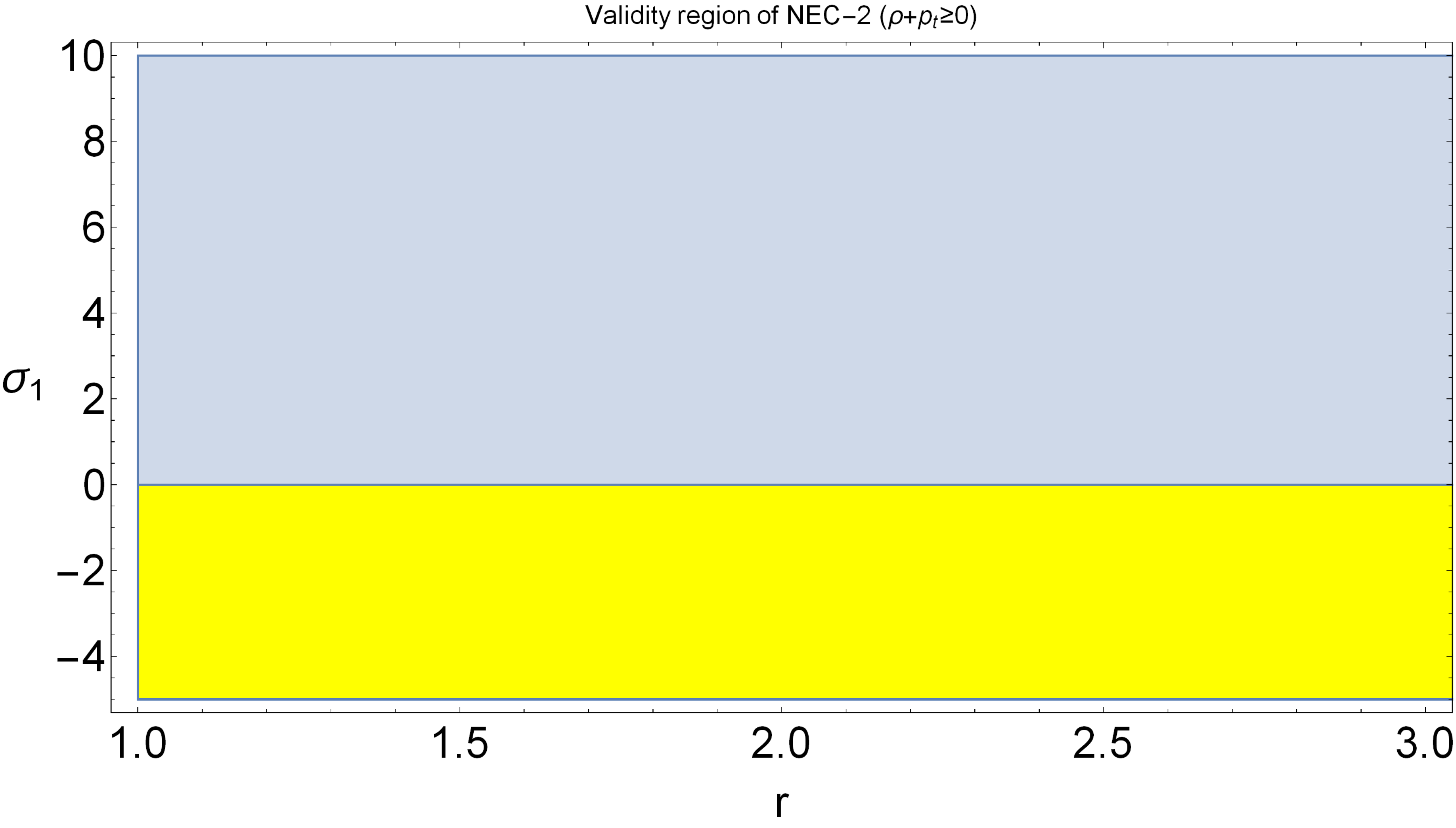}\label{fig:1b}}
\hfill
\subfloat[NEC-near the throat]{\includegraphics[width=0.5\textwidth]
{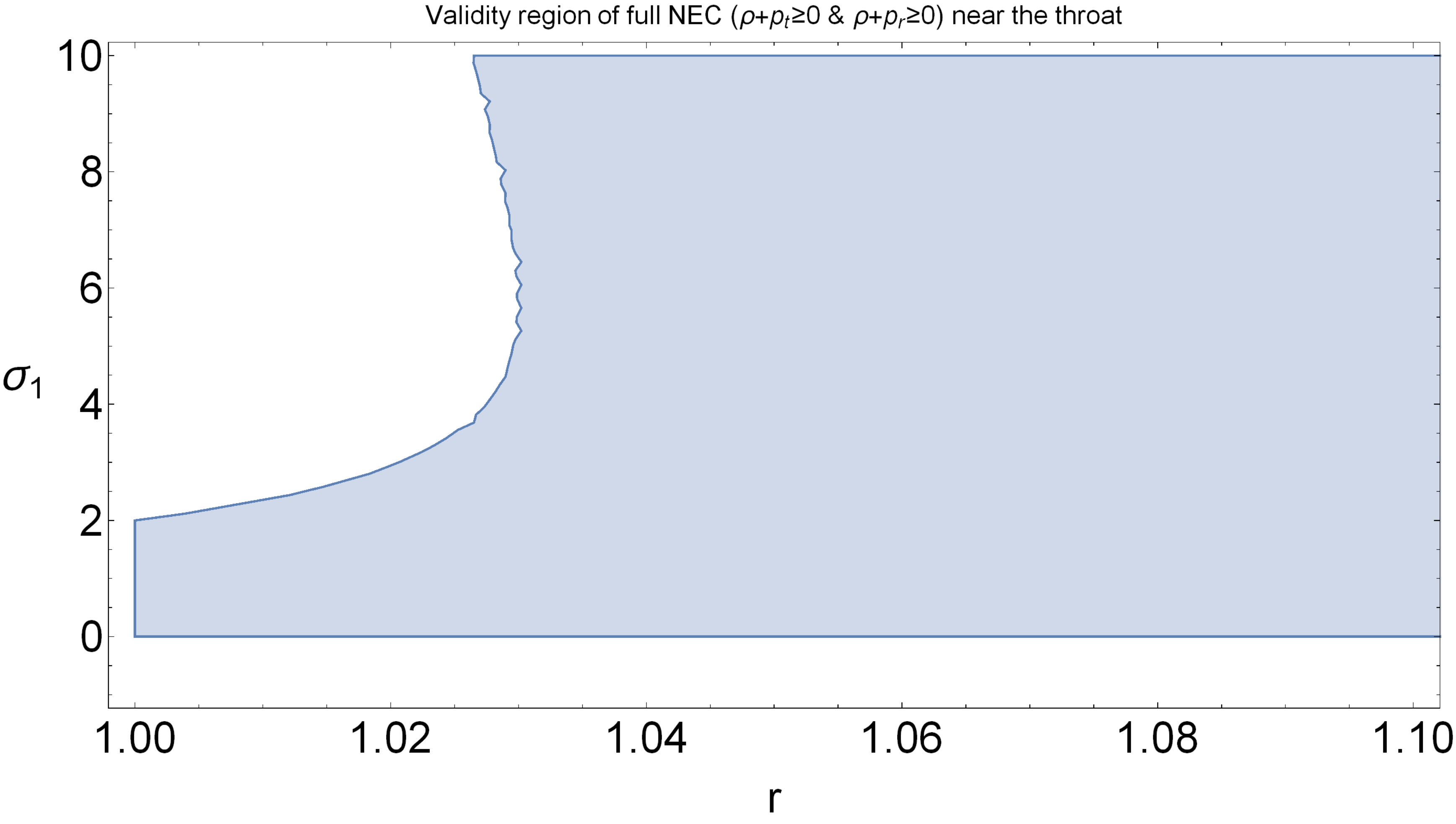}\label{fig:1c}}
\subfloat[NEC-not close to the throat]{\includegraphics[width=0.5\textwidth]
{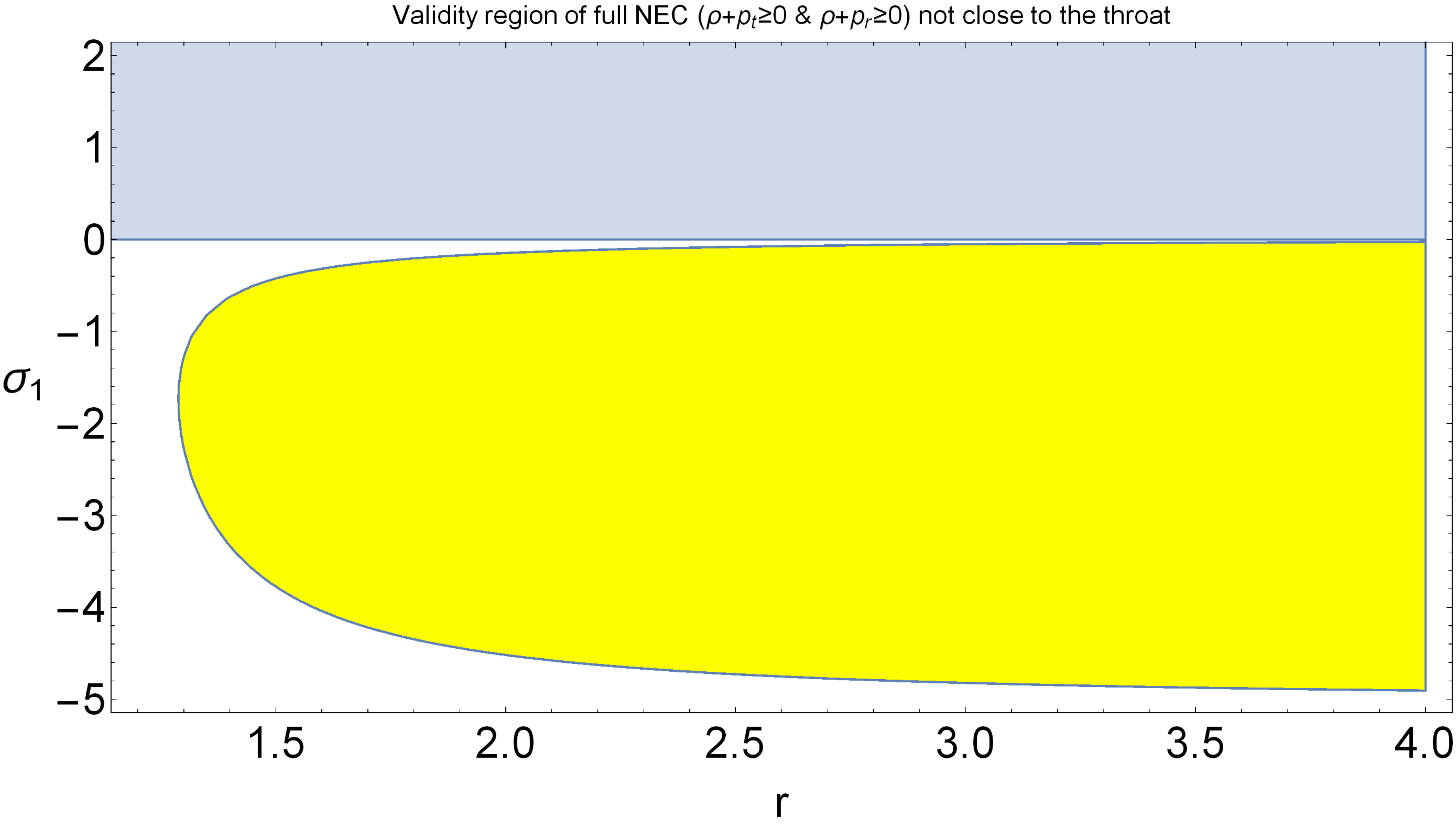}\label{fig:1d}}
\caption{\scriptsize{Validity of NEC-1 ($\rho+p_r\geq0$) given by \eqref{prpr}, NEC-2
($\rho+p_t\geq0$) given by \eqref{ptpt} and the full condition for the
validity of NEC ($\rho+p_r\geq0$ and $\rho+p_t\geq0 $)  for the generic
anisotropic fluid in Brans-Dicke theory. The yellow regions represent the
regions where $\gamma=1$ whereas the blue regions represent when $\gamma=-1$.
For these plots, we have chosen the values $r_0=\phi_0=d=\omega_0=1$ and
$\sigma_2=1$.}}
\end{figure}

\begin{figure}[H]
\captionsetup{justification=raggedright}
\subfloat[WEC-near the throat]{\includegraphics[width=0.5\textwidth]
{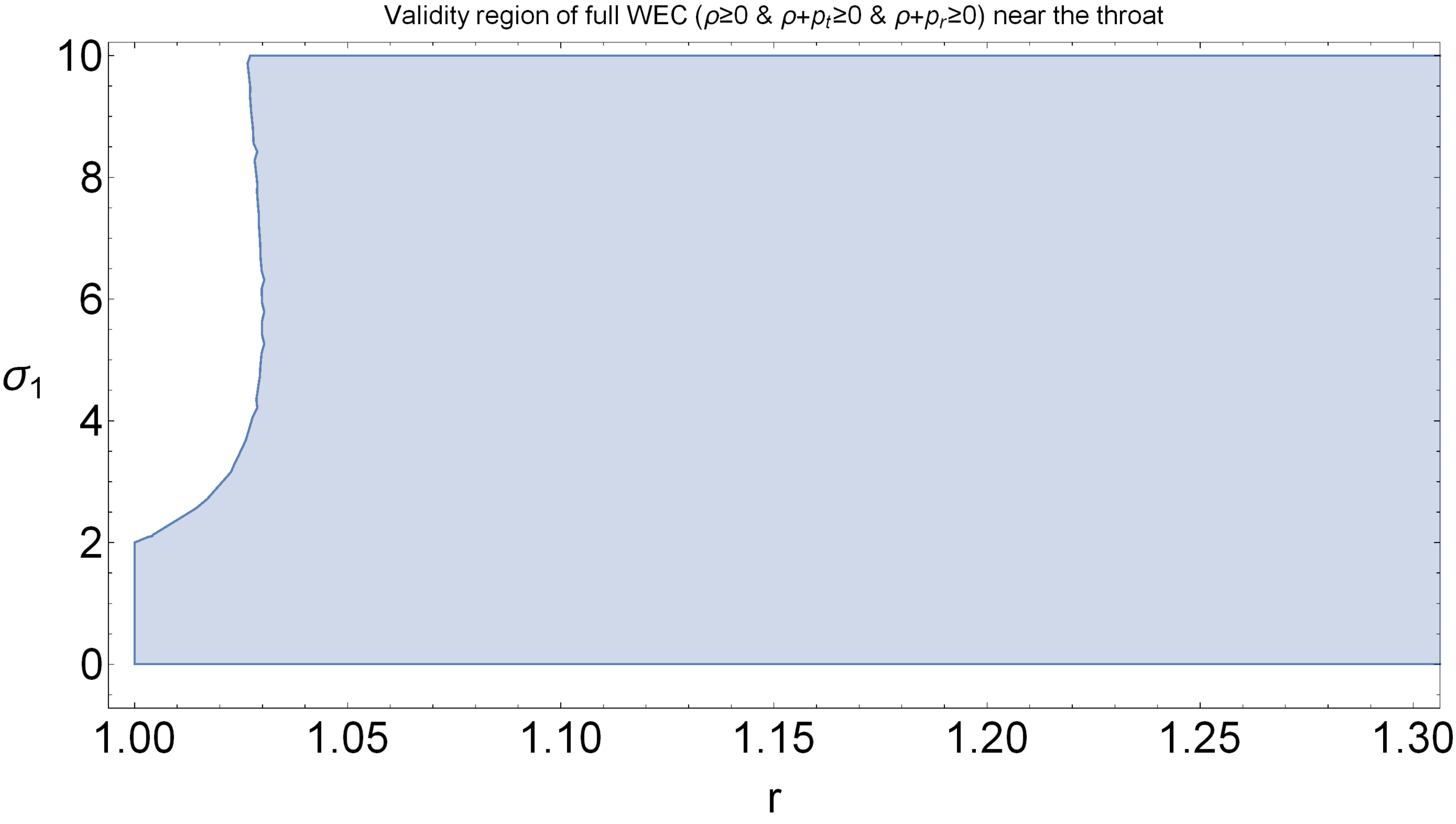}\label{fig:2a}}
\hfill
\subfloat[WEC-not close to the throat]{\includegraphics[width=0.5\textwidth]
{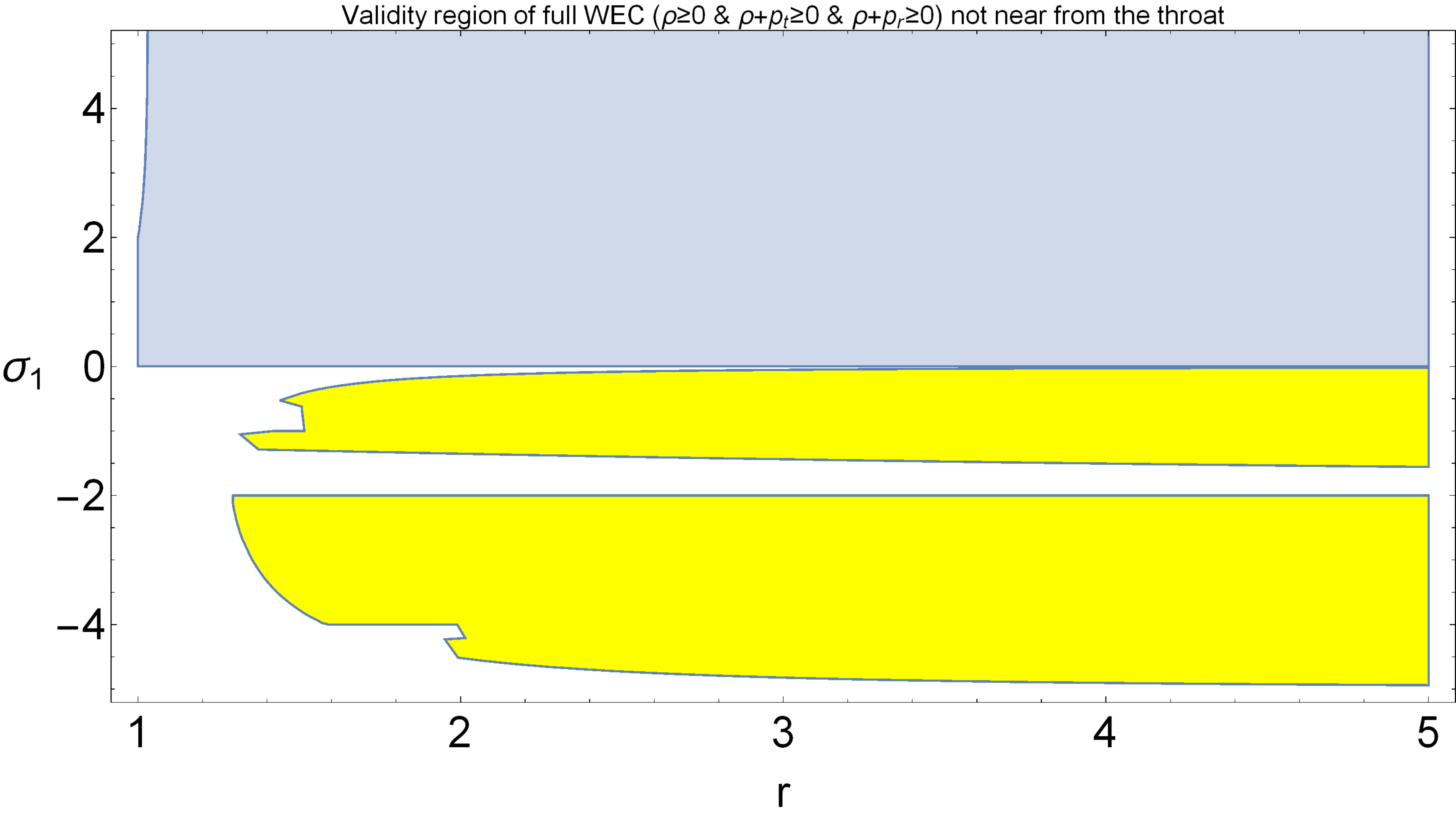}\label{fig:2b}}
\caption{\scriptsize{Validity of WEC  ($\rho\geq 0$, $\rho+p_r\geq0 $ and $\rho+p_t\geq0  $)
given by the validity of \eqref{rhorho}-\eqref{ptpt} for the generic anisotropic
fluid in Brans-Dicke theory. The figure on the right represents the validity of
WEC near the throat whereas the figure on the left shows the validity for
locations that are not close to the throat. The yellow regions represent the
regions where $\gamma=1$ whereas the blue regions represent when $\gamma=-1$.
For these plots, we have chosen the values $r_0=\phi_0=d=\omega_0=1$ and
$\sigma_2=1$.}}
\end{figure}

\begin{figure}[H]
\centering
\includegraphics[width=0.6\textwidth]{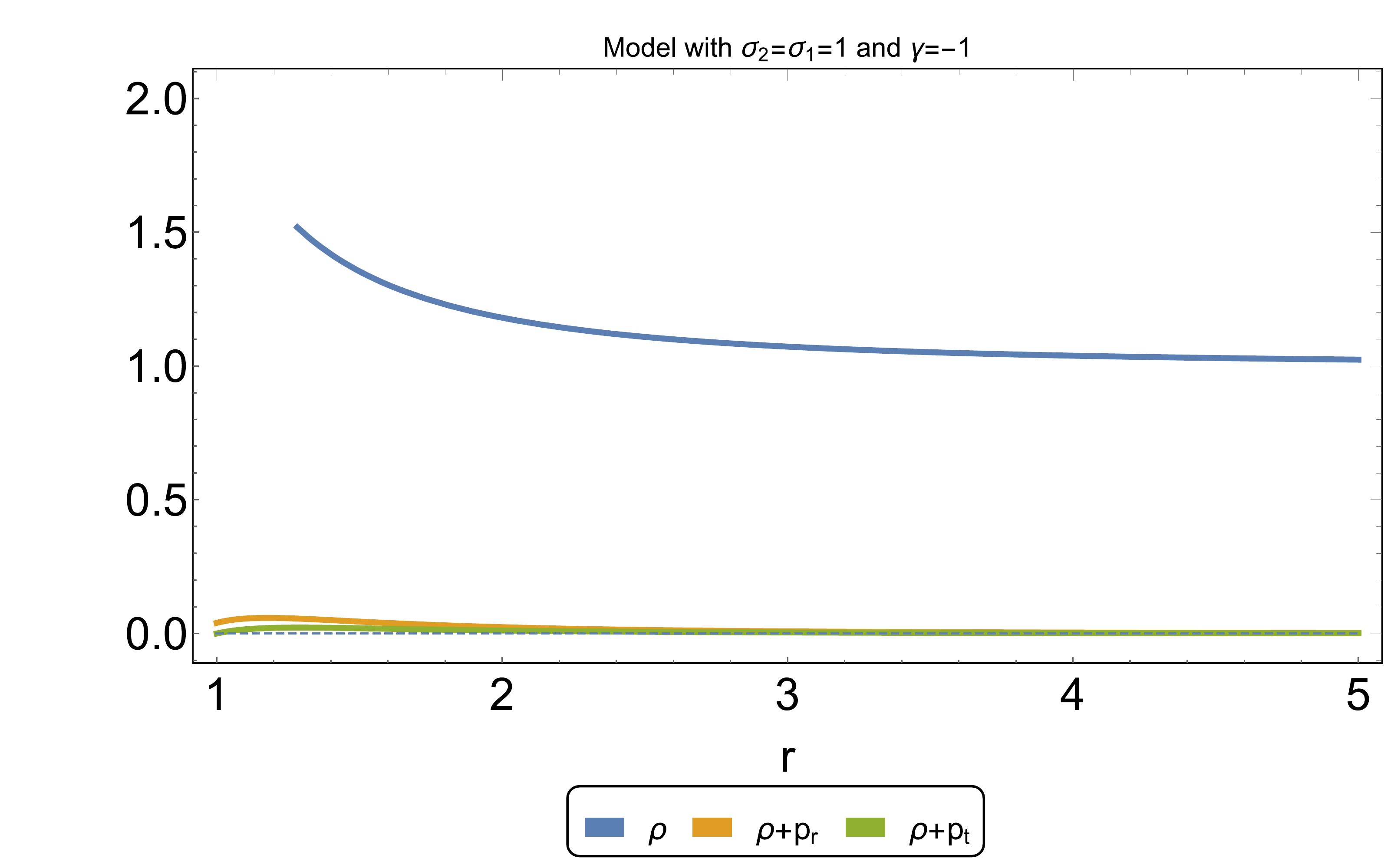}
\caption{\scriptsize{Energy density, sum of the radial pressure and the energy density and
the sum of the lateral pressure and the energy density for the generic
anisotropic fluid in Brans-Dicke theory where $\sigma_1=\sigma_2=1$ and
$\gamma=-1$. We have further chosen the values $r_0=\phi_0=d=\omega_0=1$.
For this model, the full WEC is always satisfied. }}\label{fig:3}
\end{figure}

To recover the case of Brans-Dicke theory we need to choose $n=1$ with
$m=-1$. Now, we will discuss the validity of the energy conditions for the
remaining parameters $\gamma$, $\sigma_1$ and $\sigma_2$. As we have pointed
out before, only the sign of $\gamma$ changes the physical motion of the
wormholes so we will set either $\gamma=1$ and $\gamma=-1$. Since we have two
free parameter ($\sigma_1$ and $\sigma_2$), we will make region plots to
check the validity of all the important energy conditions. For this model we
have that the corresponding energy conditions are
\begin{eqnarray}\nonumber
\rho&=&\frac{(2\gamma-1)\sigma_{1}^2}{\gamma}+\frac{8\pi\sigma_{1}(\gamma
(\sigma_{2}-3)-2\sigma_{1}(\sigma_{1}+\sigma_{2}-3))}{\gamma(\sigma_{1}
+\sigma_{2}+3)}+\frac{16\pi\sigma_{1}^3r^{\sigma_{2}+2}}{\gamma\sigma_{1}
+\gamma}+\frac{16\pi r^{\sigma_{1}+\sigma_{2}+3}}{\gamma}\\\nonumber
&&+\frac{\sigma_{1}\left(\sigma_{1}(\sigma_{1}+16\pi(3\sigma_{1}-2)+2)-2\gamma
\left(\sigma_{1}^2+5\sigma_{1}+6\right)\right)r^{\sigma_{2}+1}}{\gamma
(\sigma_{1}+2)}-\frac{16\pi r\sigma_{1}^3}{\gamma(\sigma_{1}+\sigma_{2}
+2)}\\
&&+\sigma_{1}\sigma_{2}+7\sigma_{1}-2\sigma_{2}\geq 0\,,\label{rhorho}\\
\rho+p_{r}&=&-2\sigma_{1}(\sigma_{1}+5)r^{\sigma_{2}+1}+2\sigma_{1}^2
+\sigma_{1}(\sigma_{2}+11)-2(\sigma_{2}+1)\geq 0\,,\label{prpr}\\
\rho+p_{t}&=&-6 \sigma_{1}r^{\sigma_{2}+1}+6\sigma_{1}-\sigma_{2}+1\geq 0\,.\label{ptpt}
\end{eqnarray}
We can see that it is not so easy to check the validity of the energy
conditions. Let us first study the case where $\sigma_2=1$ to visualise
better the behaviour of the energy conditions. In that case, we are able to
create 2D region plots for the validity of the energy conditions.
Fig.~\ref{fig:1} shows the validity of $\rho\geq0$ (see \eqref{rhorho}) for
different values of $\sigma_1$ and $\gamma=1$ or $\gamma=-1$. Each
blue(yellow) regions represent the validity of this condition for
$\gamma=-1$($\gamma=1$). The green regions are the intersection regions where
this condition is valid for $\gamma=1$ and $\gamma=-1$. As we can see from
the figure, there is not so much difference in the valid region for positive
or negative values of $\gamma$. However, one can directly see that for the
region where $-2\lesssim\sigma_1\lesssim -1.3$, the condition $\rho>0$  will be never
true. For other values, one can notice that the validity of this condition
depends on the location of the observer. For an observer who is far away from
the throat (located at $r_0=1$), the condition $\rho\geq0$ will be always
true. However, for an observer who is located near the throat, this condition
will be violated for some values of $\sigma_1$. Figs.~\ref{fig:1a} and
\ref{fig:1b} show similar region plots for the validity of NEC-1
($\rho+p_r\geq0$) and NEC-2 ($\rho+p_t\geq0$) given by the validity of the
inequalities \eqref{prpr} and \eqref{ptpt} respectively. For almost all
$\sigma_1$, NEC-1 depends on the location of the observer and the sign of
$\gamma$. However, there exits a region for $\gamma=-1$ given by
$-1\lesssim\sigma_1\lesssim 2$ where NEC-1 is always valid independently of
the location of the observer. For positive values of $\gamma$, it does not
exist a region where NEC-1 is valid everywhere. NEC-2 is independent of the
location of the observer. For $\gamma=-1$, NEC-2 is satisfied always if
$\sigma_1\gtrsim 0$ and for $\gamma=1$, $\sigma_1\lesssim 0$ is required.
Hence, there are not regions where NEC-1 and NEC-2 are valid for the
intersections $\gamma=1$ and $\gamma=-1$ regions. In Figs.~\ref{fig:1c}~ and~
\ref{fig:1d}~ are depicted region plots for the validity of the full NEC
energy condition ($\rho+p_r\geq0$ and $\rho+p_t\geq0 $) near the throat and
also for locations that are not so close to the throat. The full NEC
condition is satisfied if Eqs.~\eqref{prpr} and \eqref{ptpt} are true. As we
can see from the figures, for $\gamma=1$, it is not possible to find a
suitable $\sigma_1$ where the full NEC is valid at every point of the space.
Moreover, at points near the throat, NEC is always invalid for $\gamma=1$.
Although, for $\gamma=-1$, in the region $0\lesssim\sigma_1\lesssim 2$, the
full NEC is valid everywhere. Finally, Figs.~\ref{fig:2a} and \ref{fig:2b}
show the validity of the full WEC ($\rho\geq 0$, $\rho+p_r\geq0$ and
$\rho+p_t\geq0  $) close and not so close to the throat respectively. From
those figures, we can see that the full WEC is valid only for some very
special regions for $\gamma=1$ and moreover for observers closer to the
throat, it would be always invalid. This is consistent with the full NEC (see
Figs.~\ref{fig:1c} and \ref{fig:1d}) since if NEC is violated, then WEC will
be also violated. On the other hand, for $\gamma=-1$, there are different
ranges where WEC is valid but only for the range where
$0\lesssim\sigma_1\lesssim 2$, WEC is valid independently of the location of
the observer. As a consistency checking, Fig.~\ref{fig:3} shows the behaviour
of $\rho,\rho+p_r$ and $\rho+p_t$ for a special model where $\gamma=-1$ and
$\sigma_1=1$. In this model, WEC is satisfied at all locations since all the
important quantities are always positive.
\normalsize
\normalsize Let us now try to analyse the model for an arbitrary shape function parameter
$\sigma_2$. In this case, we have three parameters, namely, $\gamma$ ,
$\sigma_1$ and $\sigma_2$. As we have said before, the sign of $\gamma$ is
important but not its strength. Figs.~\ref{fig:4} show region plots for the
validity of the full NEC and WEC for positive and negatives values of
$\gamma$. One can notice that it is not possible to model wormholes
satisfying the full NEC everywhere for positive $\gamma$ since depending on
the location of the observer, that energy condition would be valid or not.
For negative values of $\gamma$, there are different models depending on
$\sigma_2$ and  $\sigma_1$ which ensures that the wormhole is supported by
non-exotic matter at every point of the space. In those specific models, the
full WEC is always satisfied.
\scriptsize{\begin{figure}[H]
\captionsetup{justification=raggedright}
\subfloat[NEC with $\gamma=1$]{\includegraphics[width=0.3\textwidth]
{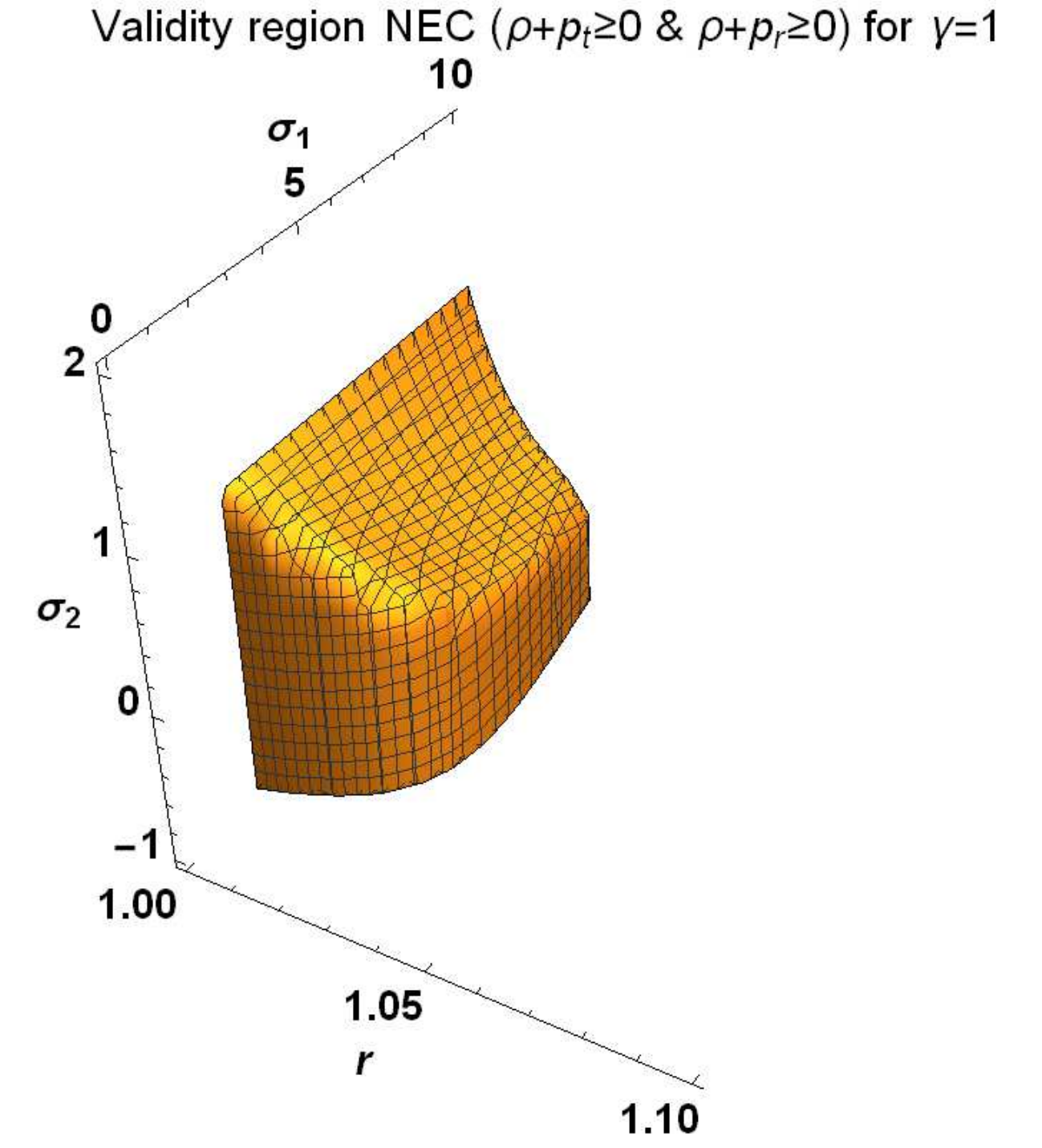}}\label{fig:4a}
\hfill
\subfloat[NEC with
$\gamma=-1$]{\includegraphics[width=0.3\textwidth]{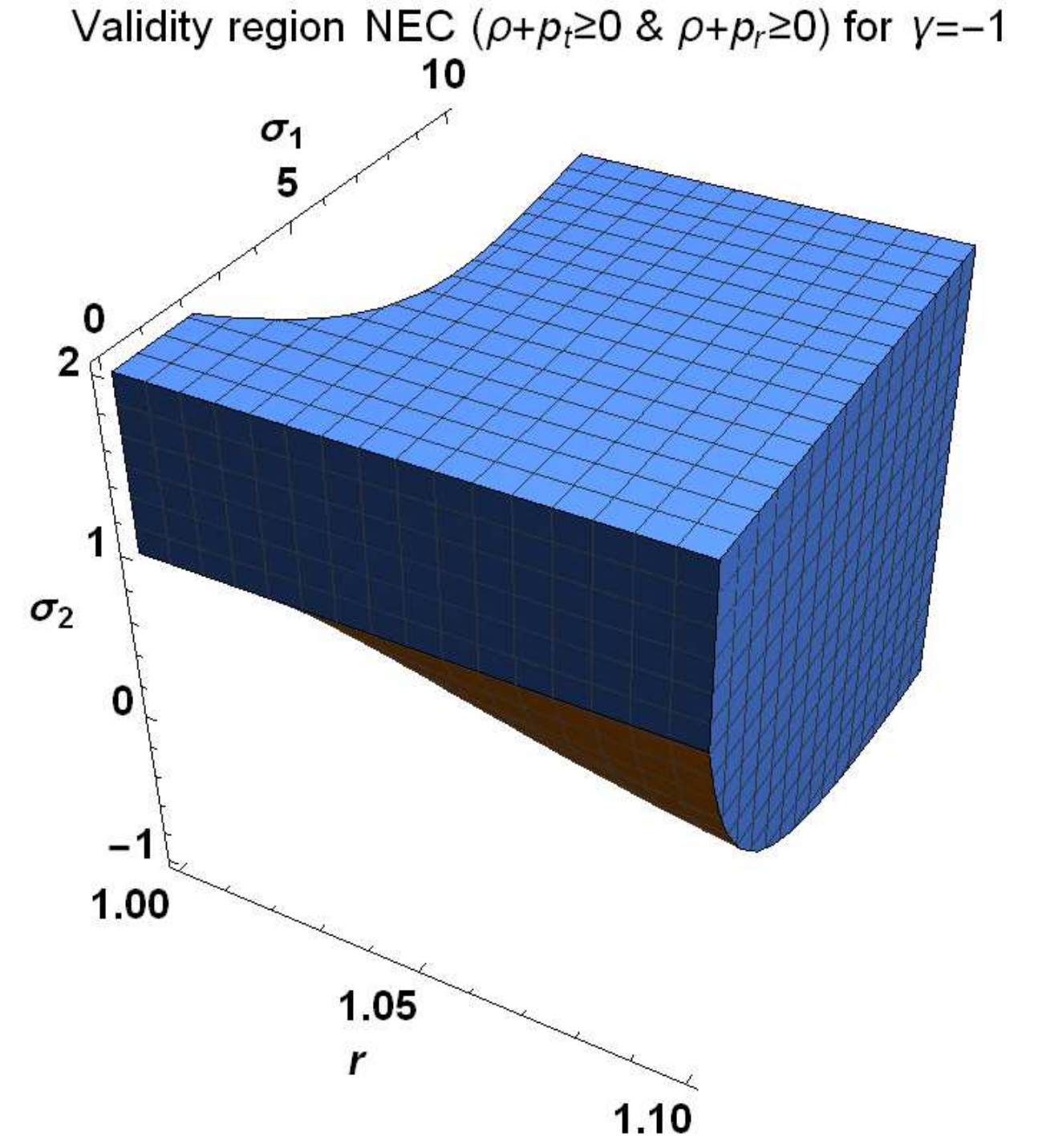}\label{fig:4b}}
\hfill
\subfloat[WEC with $\gamma=-1$]{\includegraphics[width=0.3\textwidth]
{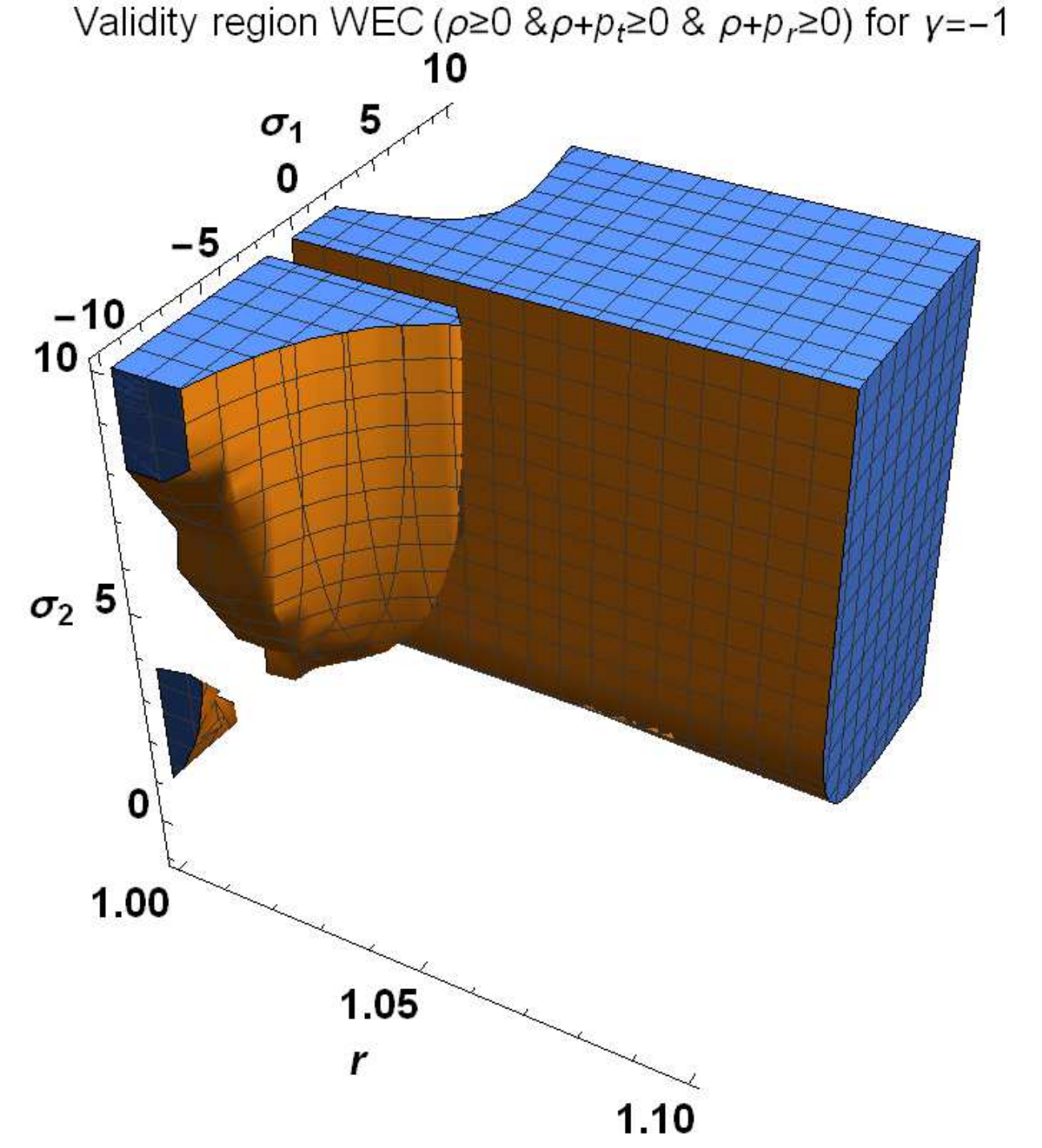}\label{fig:4c}}
\caption{\scriptsize{Validity of NEC  ($\rho+p_r\geq0$ and $\rho+p_t\geq0  $) and WEC
($\rho\geq 0$, $\rho+p_r\geq0$ and $\rho+p_t\geq0$) given by the
validity of \eqref{rhorho}-\eqref{ptpt} for the generic anisotropic fluid in
Brans-Dicke theory. The figure on the left represents the validity of NEC for
$\gamma=1$ whereas the figure on the centre represents the validity for
$\gamma=-1$. Lastly, the figure on the right shows the validity of WEC for
$\gamma=-1$. For these plots, we have chosen the values $r_0=\phi_0=d=\omega_0=1$.}}\label{fig:4}
\end{figure}}
\normalsize
\subsection{Induced gravity}\label{induced}
\begin{figure}[H]
\captionsetup{justification=raggedright}
\subfloat[NEC with
$\gamma=1$]{\includegraphics[width=0.3\textwidth]{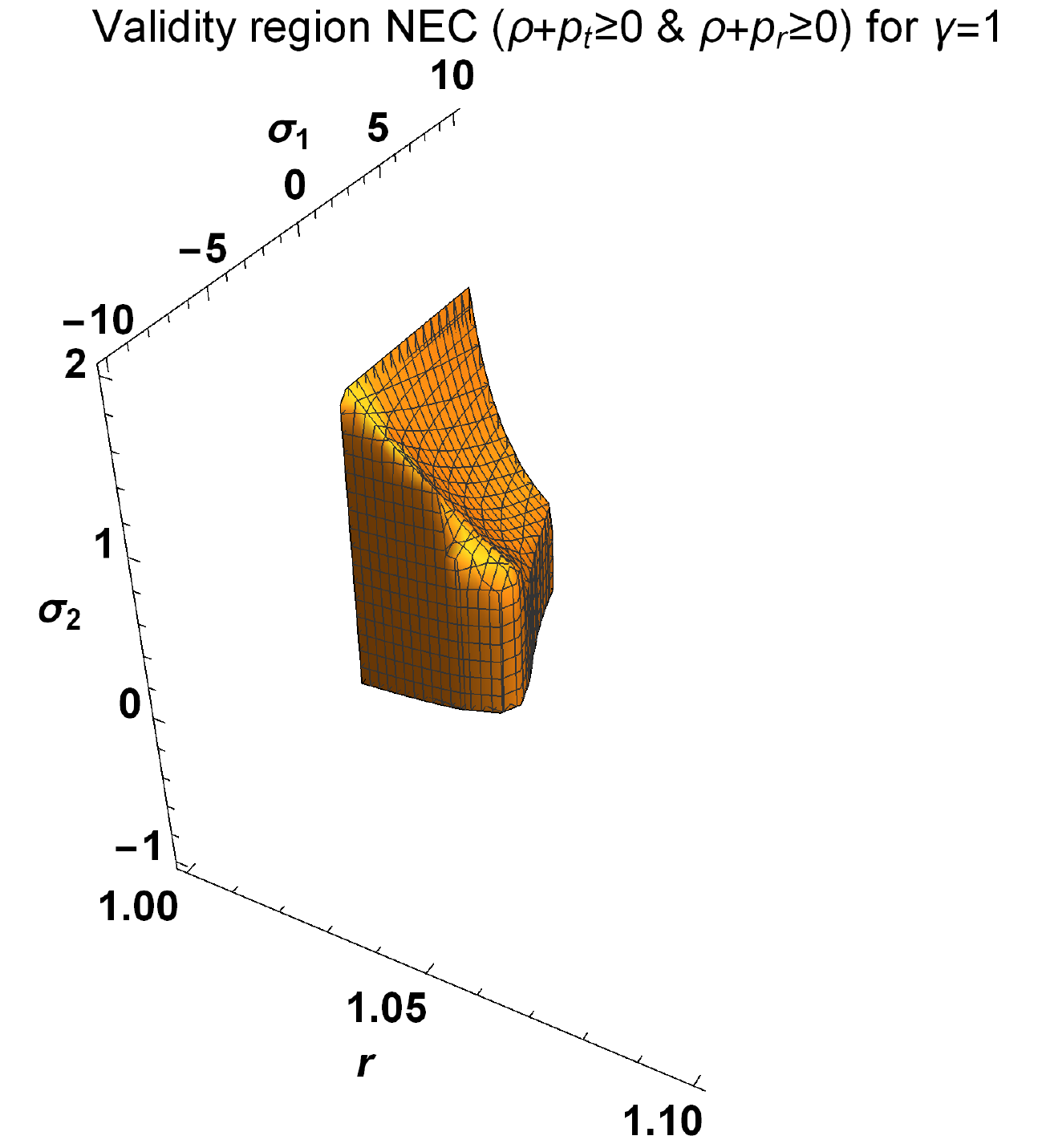}\label{fig:5a}}
\hfill
\subfloat[NEC with
$\gamma=-1$]{\includegraphics[width=0.3\textwidth]{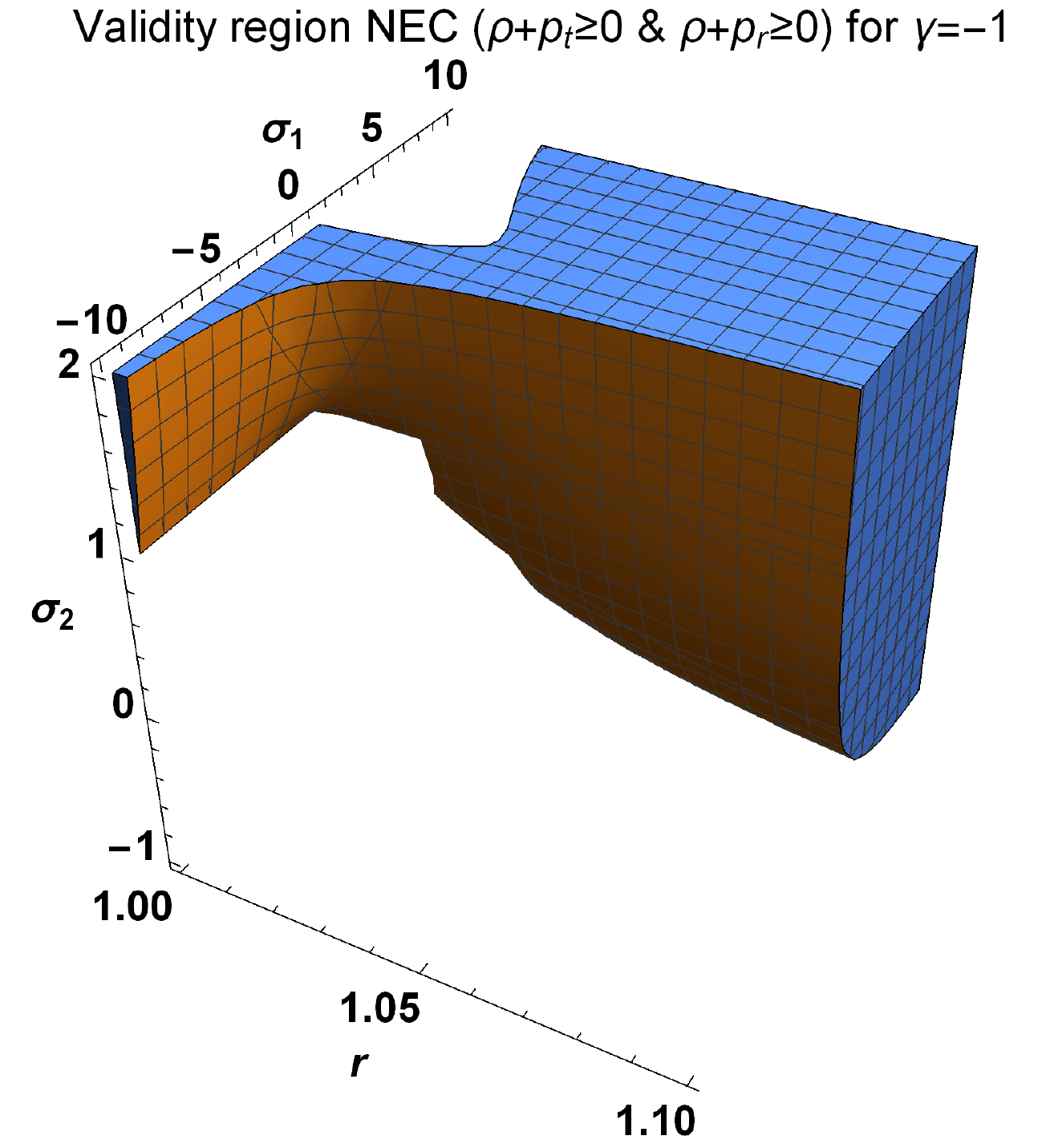}\label{fig:5b}}
\hfill
\subfloat[WEC with
$\gamma=-1$]{\includegraphics[width=0.3\textwidth]{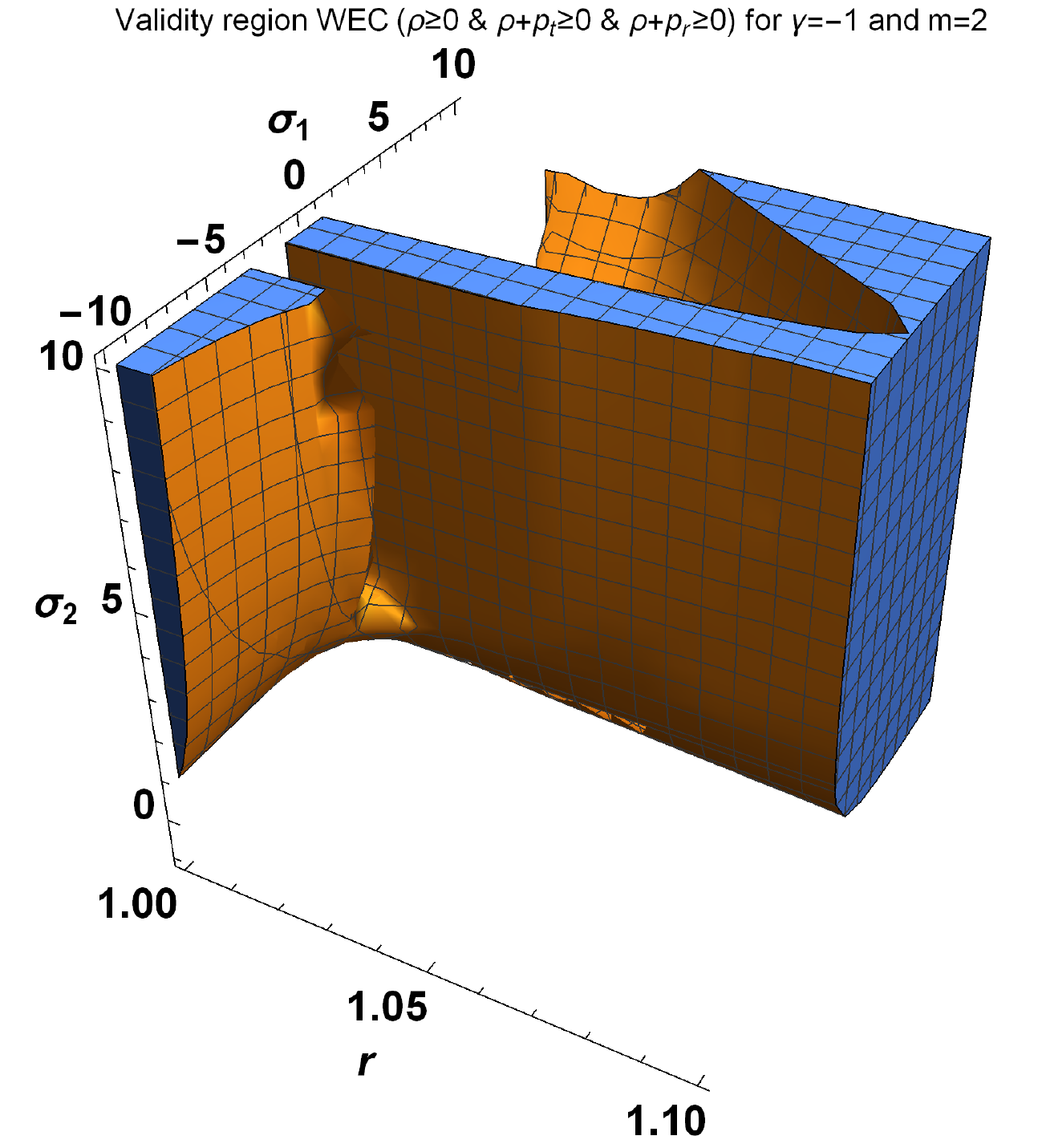}\label{fig:5c}}
\caption{Validity of NEC  ($\rho+p_r\geq0$ and $\rho+p_t\geq0  $) and WEC
($\rho\geq 0,$ $\rho+p_r\geq0$ and $\rho+p_t\geq0  $) given by the
validity of \eqref{rhorho2}-\eqref{ptpt2} for the generic anisotropic fluid
in Induced Gravity. The figure on the left represents the validity of NEC for
$\gamma=1$ whereas the figure on the centre represents the validity for
$\gamma=-1$. Lastly, the figure on the right shows the validity of WEC for
$\gamma=-1$ and $m=2$. For these plots, we have chosen the values
$r_0=\phi_0=d=\omega_0=1$. We can notice that various model exist where the
full WEC is valid for $\gamma=-1$ whereas for $\gamma=1$, it is not possible
to find that NEC is valid for every location.}\label{fig:5}
\end{figure}

\begin{figure}[H]
\centering
\includegraphics[width=0.7\textwidth]{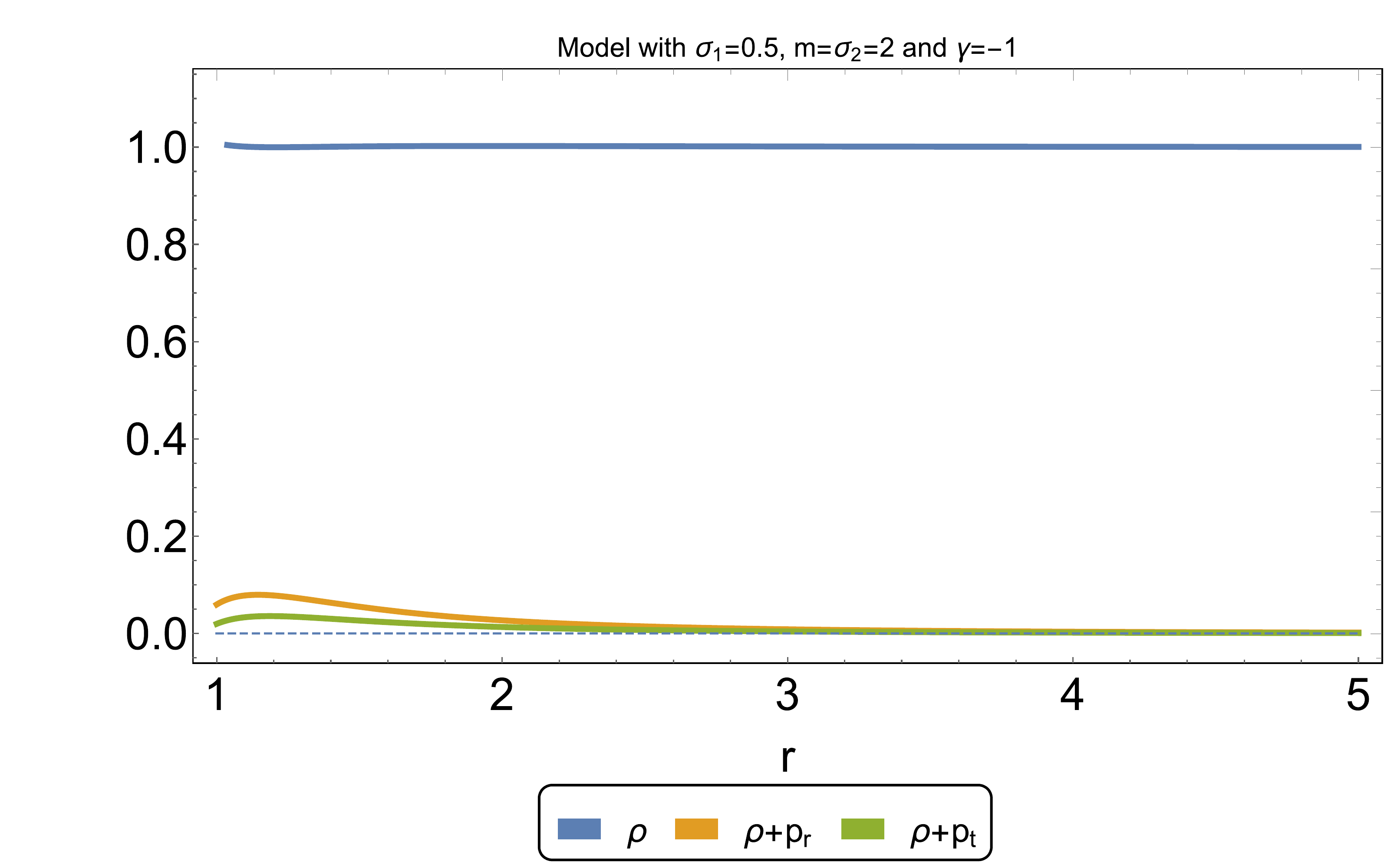}
\caption{Energy density, sum of the radial pressure and the energy density and
the sum of the lateral pressure and the energy density for the generic
anisotropic fluid in Induced Gravity  where $\sigma_1=0.5$, $\sigma_2=m=2$ and
$\gamma=-1$. We have further chosen the values $r_0=\phi_0=d=\omega_0=1$.
For this model, the full WEC is always satisfies.  }\label{fig:6}
\end{figure}

In this section, we will study the energy conditions for the Induced Gravity
case. To recover this case, we must choose $n=2$ with $m>0$. Then, we have 4
free parameters, namely $m$, $\sigma_1,\sigma_2$ and $\gamma$. Doing a
similar approach as we did in the previous section, we can distinguish
between models that do not violate the energy conditions. Without going into
too much details as in the previous section, in this section we will only
show the validity of the full NEC and full WEC. If WEC is valid, all the
other energy conditions will be valid too. The validity of WEC will be true
if all the following three inequalities hold,
\begin{eqnarray}
\rho&=&-\frac{m \sigma_{1}^3 r^{-(m+2) \sigma_{1}-1}}{(m+2) \sigma_{1}+1}
+\frac{1}{16} \sigma_{1}^2 r^{-(m+2) \sigma_{1}-2} \left(\frac{16 m \sigma_{1}
r^{-\sigma_{2}}}{(m+2)\sigma_{1}+\sigma_{2}+2}-\frac{16 ((m-2) \sigma_{1}+2)}
{(m+2) \sigma_{1}+2}+\frac{1}{\pi}\right)\nonumber\\
&&+\frac{\sigma_{1}^2 (16 \pi  (m \sigma_{1}-\sigma_{2}+3)-(m+2) \sigma_{1}
-\sigma_{2}-3) r^{-(m+2) \sigma_{1}-\sigma_{2}-3}}{16 \pi  ((m+2) \sigma_{1}
+\sigma_{2}+3)}\nonumber\\
&&+\frac{\gamma  \left(8 \sigma_{1}^3+\sigma_{1}^2 (6 \sigma_{2}+26)+\sigma_{1}
\left(\sigma_{2}^2+8 \sigma_{2}+8 \pi  (\sigma_{2}-3)+21\right)-\sigma_{2}
(\sigma_{2}+3)\right) r^{-2 \sigma_{1}-\sigma_{2}-3}}{8 \pi  (2 \sigma_{1}
+\sigma_{2}+3)}\nonumber\\
&&-\frac{\gamma  \sigma_{1} (2 \sigma_{1}+3) r^{-2 \sigma_{1}-2}}{4 \pi }
+1\geq 0\,,\label{rhorho2}\\
\rho+p_r&=&\frac{\gamma  \left(4 \sigma_{1}^2+\sigma_{1} (\sigma_{2}+11)
-\sigma_{2}-1\right) r^{-2\sigma_{1}-\sigma_{2}-3}}{8 \pi }-\frac{\gamma
\sigma_{1} (2 \sigma_{1}+5) r^{-2 \sigma_{1}-2}}{4 \pi}\geq 0\,,\label{prpr2}\\
\rho+p_t&=&\frac{\gamma  (12 \sigma_{1}-\sigma_{2}+1) r^{-2 \sigma_{1}
-\sigma_{2}-3}}{16 \pi }-\frac{3\gamma  \sigma_{1} r^{-2 \sigma_{1}-2}}{4 \pi}
\geq0\label{ptpt2}\,.
\end{eqnarray}
Note that the validity of the last two inequalities do not depend on the
parameter $m$. Hence, the validity of the full NEC will not depend on the
parameter $m$. Figs.~\ref{fig:5a} and \ref{fig:5b} show the validity of NEC
for $\gamma=1$ and $\gamma=-1$ respectively. Exactly as the Brans-Dicke case,
NEC cannot be true for every location when $\gamma$ is positive. Moreover,
the problem comes near the throat. Hence, the full WEC will be also not true
at every location for Induced Gravity when $\gamma$ is positive. On the other
hand, for negative $\gamma$, it is possible to ensure the validity of NEC for
some parameters $\sigma_1$ and $\sigma_2$. Fig.~\ref{fig:5c} shows the
validity of WEC for $m=2$ and $\gamma=-1$. Since $\rho$ depends on $m$, the
validity of WEC will depend on $m$ too. For bigger values of $m$, the
validity of WEC is more constraint. However, it always exists a small range
of values of $\sigma_1$ and $\sigma_2$  where WEC will be true at every
location (even near the throat). From the figure one can notice that this
small region is $-1\lesssim\sigma_1\lesssim 1$. Fig.~\ref{fig:6} depicts the
energy density and the sum of the pressures with the energy density for a
model in this range, where $\sigma_1=0.5$. In the latter figure, we have
further chosen $m=\sigma_2=2$ and $\gamma=-1$. One can see from the figure,
that WEC is always true in this model.

\section{Isotropic Fluid ($p_r=p_t=p$)}\label{sec44}
In this section, we will study the isotropic fluid case when $p_r=p_t=p$. By choosing this kind of fluid and substituting $e^{-b(r)}=1-
\frac{\beta(r)}{r}$ in the field equations \eqref{2.12}-\eqref{2.14}, we find the following differential equation for the shape function,
\begin{eqnarray}\label{2.23}
\nonumber&&\frac{\gamma}{dr\kappa^2}{\phi_0}^{n}\left(\frac{d}{r}\right)^{n
\sigma_1}\bigg[d^2r(n\sigma_1-1)\beta'(r)+\left\{-2n{\sigma_1}^3(\sigma_1-1)
+d^2\left(3-7n\sigma_1+2n{\sigma_1}^2-2n^2{\sigma_1}^2\right)\right\}\beta(r)\\
&&+2nr\sigma_1\left\{{\sigma_1}^2(\sigma_1-1)+d^2(3+n\sigma_1-\sigma_1)\right\}
\bigg]=0\,.
\end{eqnarray}
We can easily solve this equation analytically giving us the following shape function
\begin{equation}\label{2.24}
\beta(r)=-\xi r+c_1 r^{-\eta}\,,
\end{equation}
where $c_1$ is an integration constant and for simplicity we have introduced the constants $A_1=d^2 (n\sigma_1-1)$,
$A_2=-2n{\sigma_1}^3(\sigma_1-1)+d^2\left\{3-7n\sigma_1+2n{\sigma_1}^2-2n^2
{\sigma_1}^2\right\}$, $A_3=2n\sigma_1\left\{{\sigma_1}^2(\sigma_1-1)+d^2
\left(3+n\sigma_1-\sigma_1\right)\right\}$, $\xi=\frac{A_3}{A_1+A_2}$ and
$\eta=\frac{A_2}{A_1}$. At the throat $r=r_0$, we have the condition $\beta(r_0)=r_0$. Using this relation, it is easily to get that the throat is located at $r_0=\left(\frac{c_1}{1+\xi}\right)^{\frac{1}{\eta+1}}$. Using the flaring-out condition at the throat, $\beta'(r_0)<1$, one notices that the condition $\eta>-1$ must be satisfied.

By replacing a power-law model described by (\ref{2.11}) and the above form of the shape function into the modified Klein-Gordon equation (\ref{2.2*}), one can obtain the potential yielding
\begin{eqnarray}\label{2.24*}
\nonumber V(\phi)&=&\frac{2nc_1\eta\gamma\sigma_1{d}^{-\eta-3}
{\phi_0}^{\frac{-\eta-3}{\sigma_1}}}{n\sigma_1+\eta+3}\phi^{\frac{n\sigma_1
+\eta+3}{\sigma_1}}+\frac{\omega_0{\sigma_1}^2(1+\xi)(-2+2\sigma_1+m\sigma_1)
d^{-2}{\phi_0}^{-2/\sigma_1}}{m\sigma_1+2\sigma_1+2}{\phi}^{\frac{m\sigma_1
+2\sigma_1+2}{\sigma_1}}\\
\nonumber&+&\frac{2n\gamma\xi\sigma_1d^{-2}{\phi_0}^{-2/\sigma_1}}{n\sigma_1+2}
{\phi}^{\frac{n\sigma_1+2}{\sigma_1}}-\frac{\omega_0c_1{\sigma_1}^2(-1+\eta
+2\sigma_1+m\sigma_1)d^{-3-\eta}{\phi_0}^{\frac{-3-\eta}{\sigma_1}}}{m\sigma_1
+2\sigma_1+\eta+3}\phi^{m+2+\frac{\eta+3}{\sigma_1}}+c_0\,,
\end{eqnarray}
where $c_0$ is an integration constant. In the latter, we will study the Brans-Dicke and Induced Gravity cases to analyse the regions when the energy conditions are valid.
\subsubsection{Brans-Dicke theory}
\begin{figure}[H]
\centering
{\includegraphics[width=0.7\textwidth]{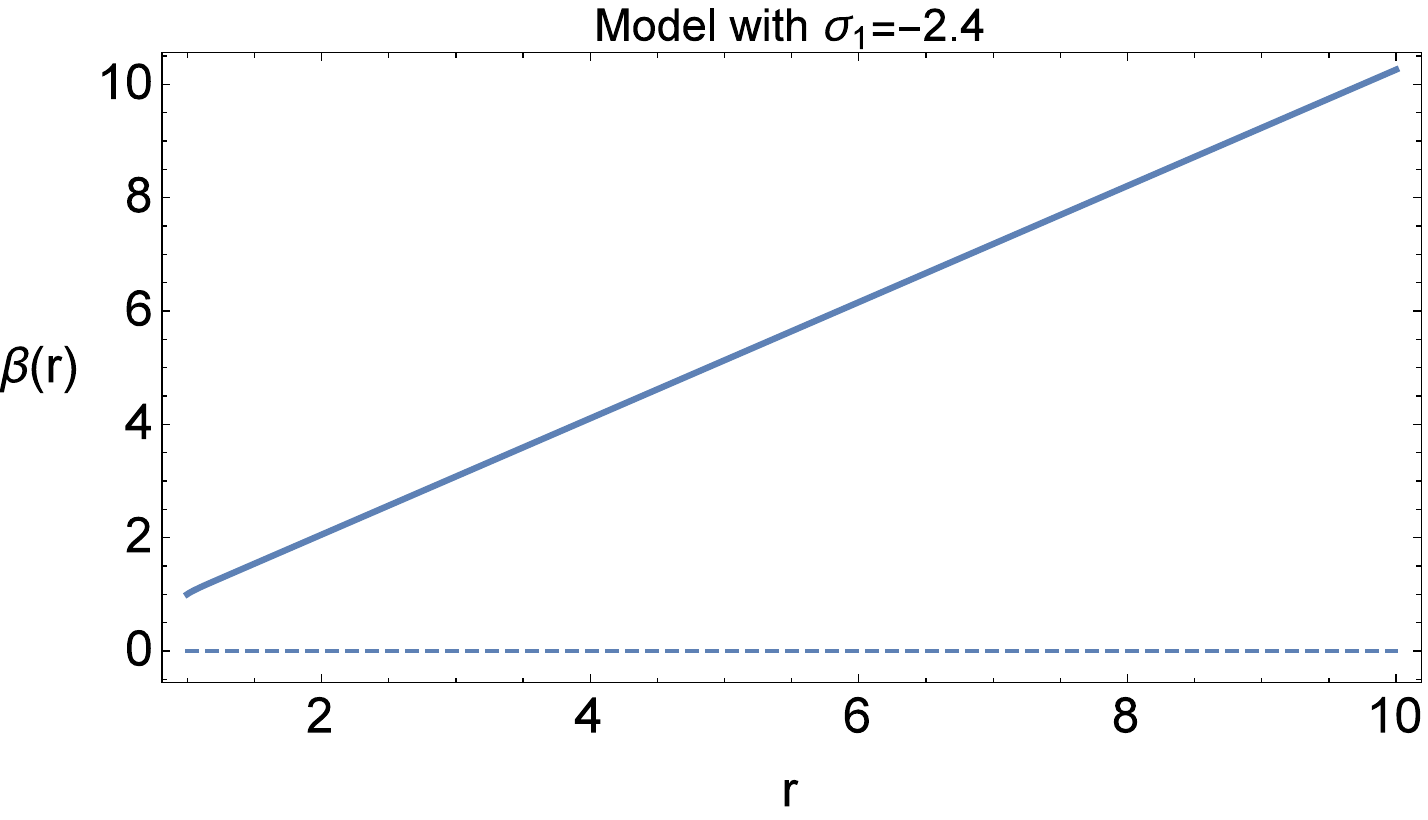}}
\hfill
\caption{The behavior of $\beta(r)$ versus $r$ taking $n=1$, $m=-1$, $\sigma_1=-2.4$, $r_0=1$ in case of BD theory.}\label{fig7}
\centering
{\includegraphics[width=0.6\textwidth]{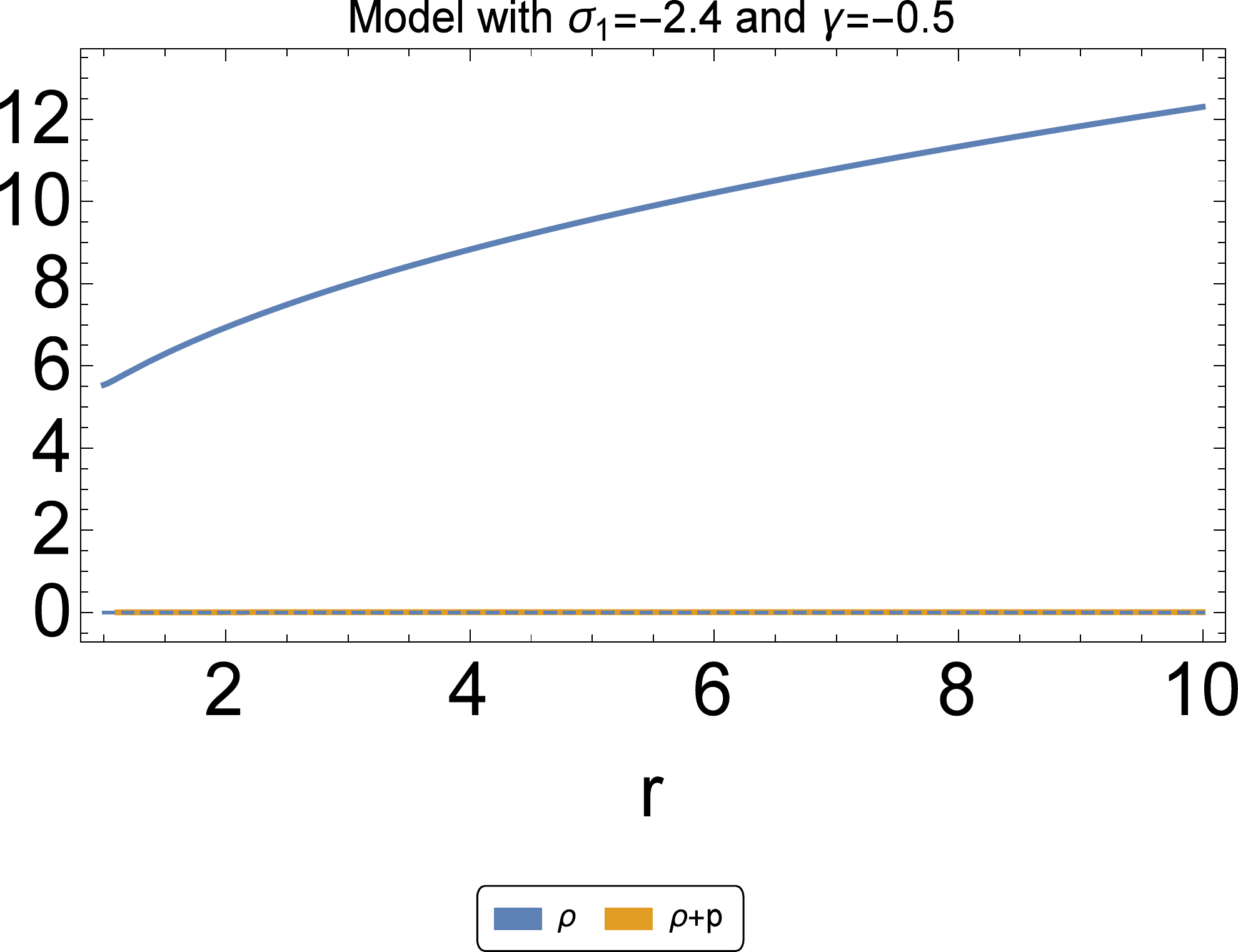}}
\caption{The behavior of $\rho$ and $\rho+p_r$ versus $r$ taking $n=1$, $m=-1$, $\sigma_1=-2.4$, $r_0=1$ for BD theory.}\label{fig8}
\end{figure}

Taking $n=1$ and $m=-1$, we get $f(R,\phi)=\gamma R \phi$ which describes Brans-Dicke theory. We will discuss the behavior of $\beta(r)$, $\rho$ and $\rho+p$ by taking the special case where the parameters $d=\omega_0=c_0=\phi_0=1$, $\sigma_1=-2.4$, $r_0=1$ and
$\gamma=-0.5$. The behavior of the shape function is shown in Fig.~\ref{fig7}. This figure shows that $\beta(r)$ is increasing and also satisfy $\beta(r)<r$. The behavior of NEC and WEC are shown in Fig.~\ref{fig8}. In that case, $\rho>0$ and $\rho+p>0$ are satisfied throughout the evolution. Then, all the energy conditions are satisfied for the parameter chosen. Thus, isotropic fluids satisfying the energy conditions can support wormholes in Brans-Dicke theory.
\subsubsection{Induced gravity}
\begin{figure}[H]
\centering
{\includegraphics[width=0.7\textwidth]{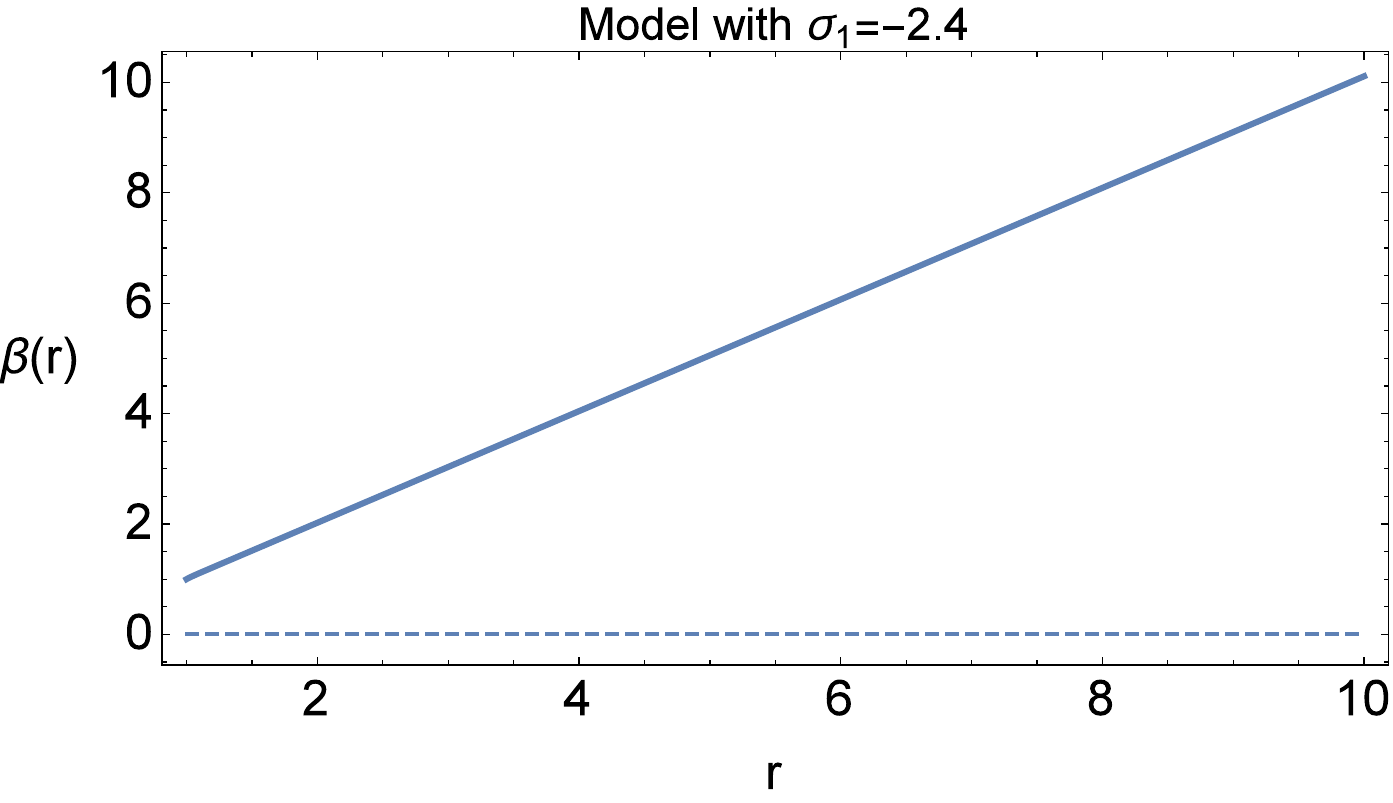}}
\caption{The behavior of $\beta(r)$ versus $r$ taking
$n=2$, $m=2$, $\sigma_1=-2.4$, $r_0=1$ in the case of the Induced gravity.}\label{fig9}
\centering
{\includegraphics[width=0.8\textwidth]{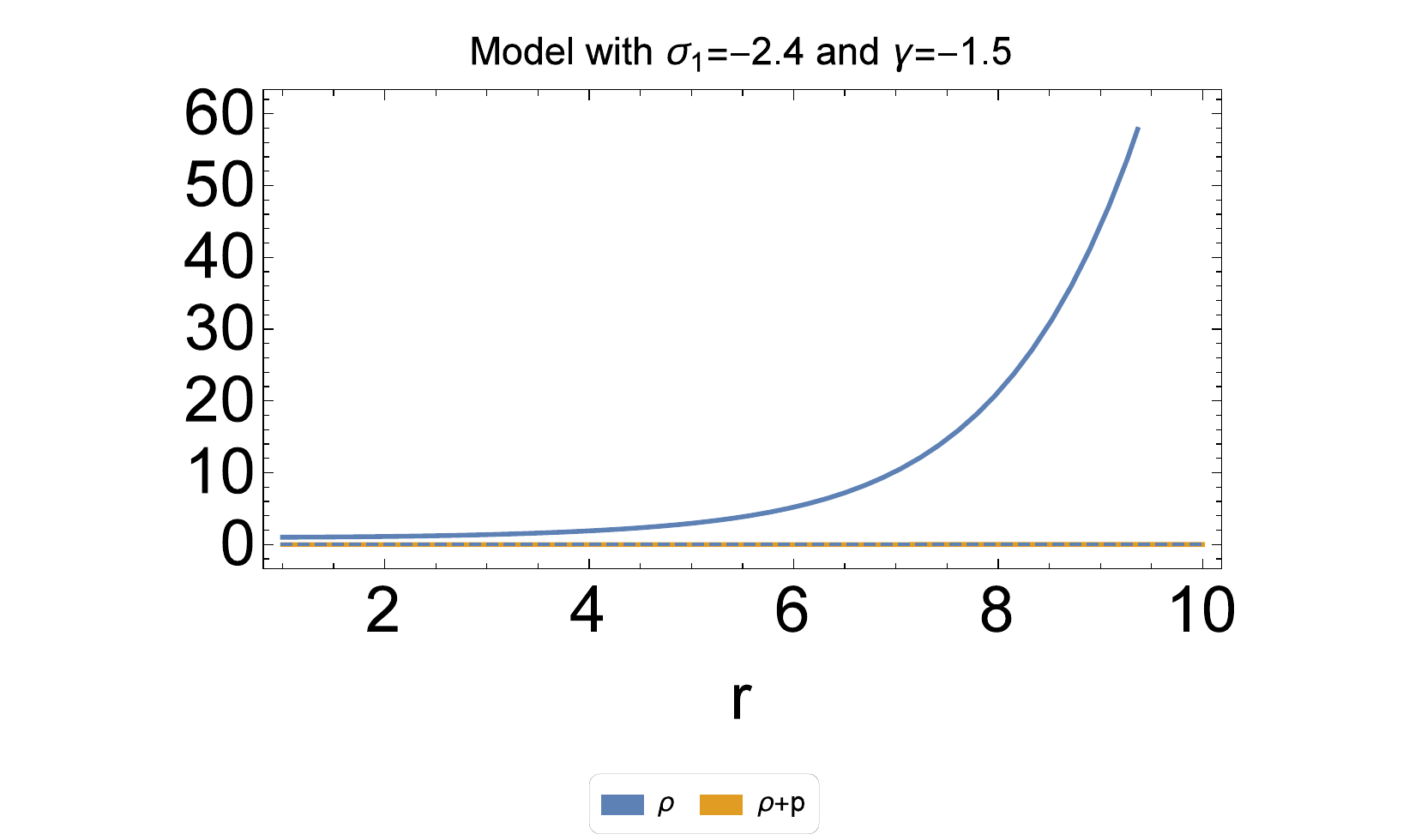}}
\caption{Plot shows the evolution of $\rho$ and $\rho+p$ in case of Induced gravity for the parameters $n=2$, $m=2$, $\sigma_1=-2.4$, $r_0=1$, $\gamma=-0.5$.}\label{fig10}
\end{figure}

Let us now explore Induced Gravity, where one needs to take $n=2$ and $m>0$. To analyse the properties of the wormhole in this theory, let us chose the parameters $\omega_0=-0.5$, $r_0=d=c_0=1$, $\phi_0=0.05$, $\sigma_1=-2.4$ and $\gamma=-0.5$.
Fig.~\ref{fig9} shows the increasing behavior of shape function
and validate the term $\beta(r)<r$. The behavior of NEC and WEC are shown in Fig.~\ref{fig10}, which shows the validity of the energy conditions throughout the evolution. Then, exactly as in Brans-Dicke theory, in Induced gravity is possible to construct wormholes supported by an isotropic fluid satisfying the energy conditions.

\section{Barotropic fluid with EoS $p_r=W(r)\rho$}\label{sec55}
In this section, we will choose a generic varying barotropic fluid with an EoS $p_r=W(r)\rho$ which involves radial pressure, energy
density and a positive radial function $W(r)$. In \cite{24a*}, Rahaman et al. have used that type of EoS with a varying parameter. For this section, we will also assume a potential of the form
\begin{equation}\nonumber
V(\phi)=\frac{V_0}{\phi^\alpha}\,.
\end{equation}
Here $V_0$ is a constant. In the following, different forms of the function $W(r)$ will be adopted to analyse different barotropic fluids which can support wormholes.
\subsection{$W(r)=W=Constant$}
First we take $W(r)=W=constant$ which is a standard barotropic fluid. By replacing this form of EoS in the field
equations (\ref{2.12})-(\ref{2.14}), we obtain the following constraint,
\begin{eqnarray}\label{2.25}
\nonumber&&\frac{1}{r\kappa^2}\left(\frac{d}{r}\right)^{-\alpha\sigma_1}{\phi_0}^{-\alpha}
\bigg[-2\kappa^2r^3V_0(W+1)+2nr\gamma\sigma_1(-2+3W+n\sigma_1W){\phi_0}^{n+\alpha}
\left(\frac{d}{r}\right)^{\sigma_1(n+\alpha)}\\
\nonumber&&-r\omega_0{\sigma_1}^2(W+1){\phi_0}^{m+2+\alpha}\left(\frac{d}{r}\right)^{\sigma_1(m+2+\alpha)}
+\omega_0{\sigma_1}^2(W+1){\phi_0}^{m+2+\alpha}\left(\frac{d}{r}\right)^{\sigma_1(m+2+\alpha)}
\beta(r)\\
\nonumber&&-\gamma{\phi_0}^{n+\alpha}\left(\frac{d}{r}
\right)^{\sigma_1(n+\alpha)}\left\{(2-4n\sigma_1+7n\sigma_1W+2n^2{\sigma_1}^2W)\beta(r)
+rW(2-n\sigma_1)\beta'(r)\right\}\bigg]=0\,.
\end{eqnarray}
The above equation cannot be easily solved analytically, so that we will explore some numerical interesting cases to study it.
\subsubsection{Brans-Dicke theory}
\begin{figure}[H]
	\captionsetup{justification=raggedright}
	\subfloat[Plot of
$\beta(r)$]{\includegraphics[width=0.5\textwidth]{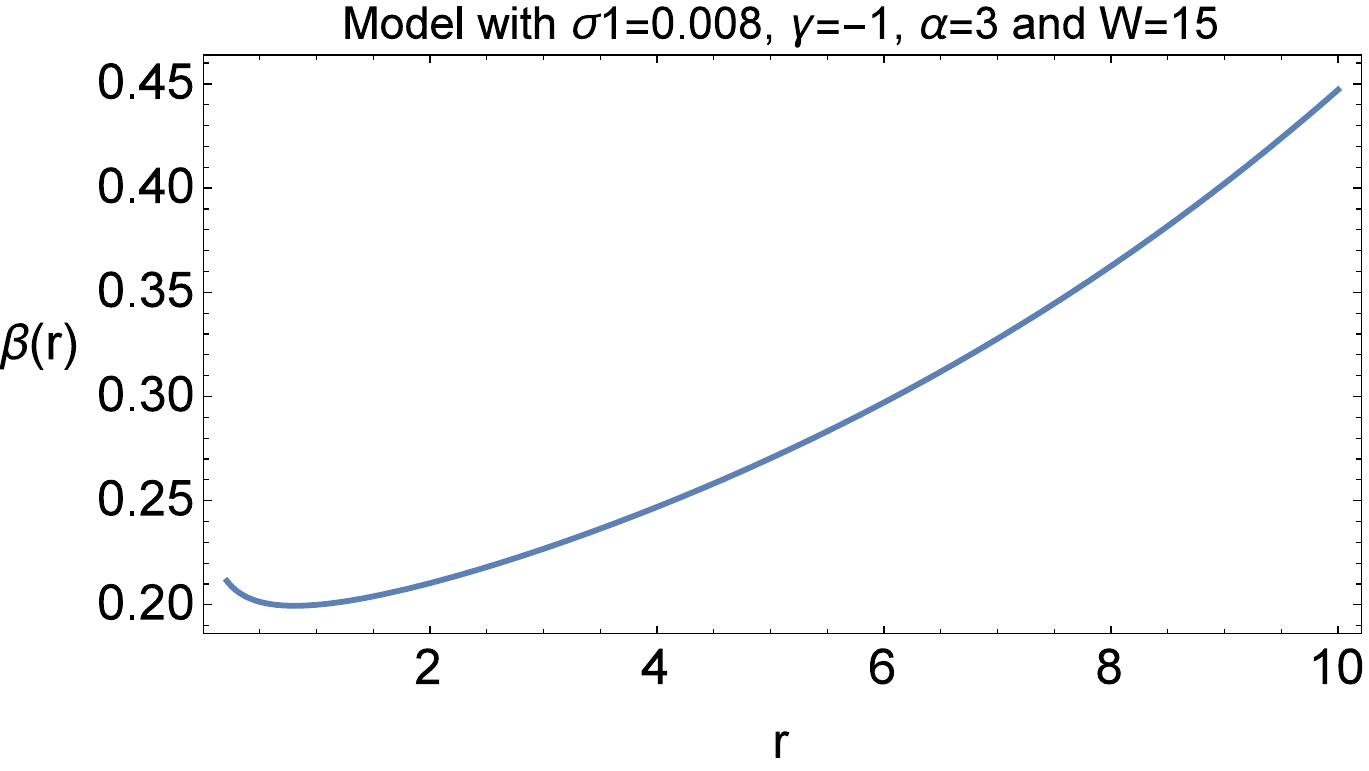}\label{fig11a}}
	\hfill
	\subfloat[Plot of
$\beta'(r)$]{\includegraphics[width=0.5\textwidth]{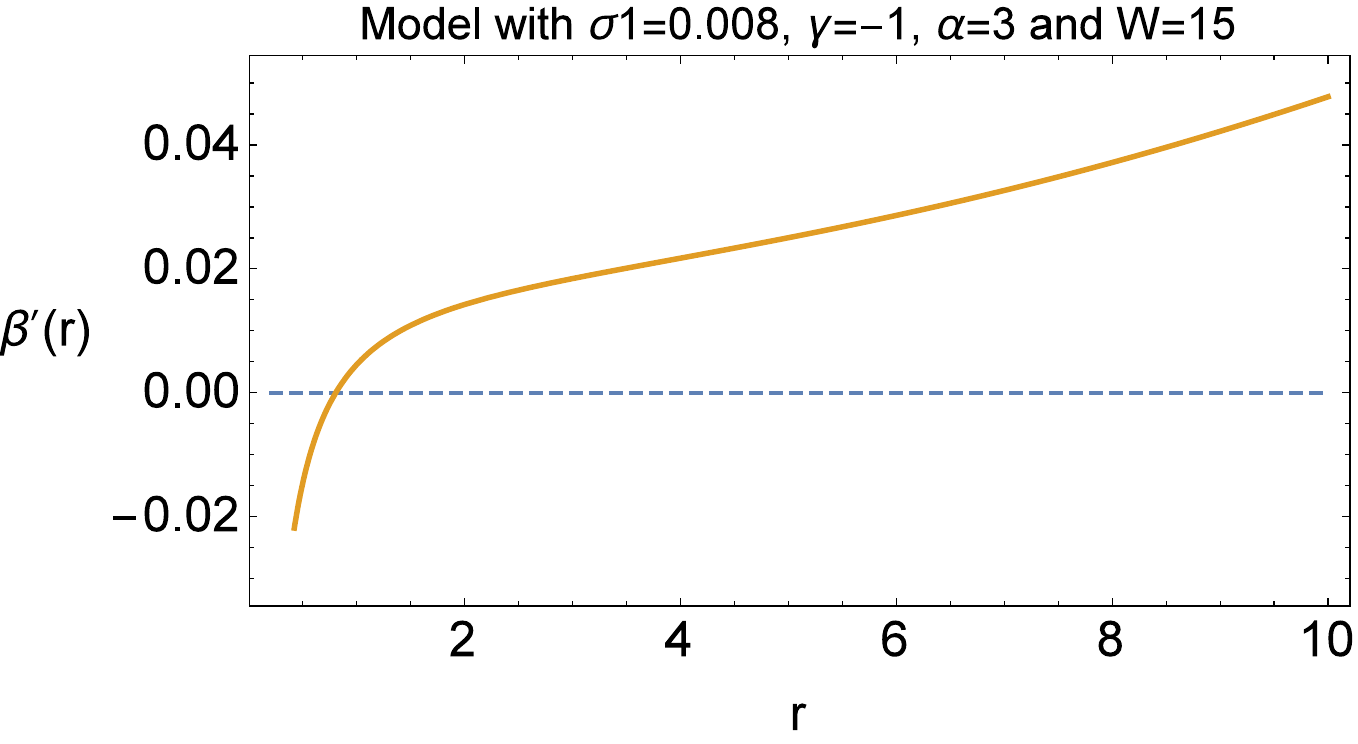}\label{fig11b}}
	\hfill
\caption{The behavior of $\beta(r)$ and $\beta'(r)$ versus $r$ taking $\omega_0=-2$, $\sigma_1=0.008$, $\gamma=-1$, $\alpha=3$, $W=15$, $\phi_0=10$, $V_0=0.1$.}\label{fig11}
\end{figure}

\begin{figure}[H]
	\captionsetup{justification=raggedright}
	\subfloat[Plot of
$\frac{\beta(r)}{r}$]{\includegraphics[width=0.5\textwidth]{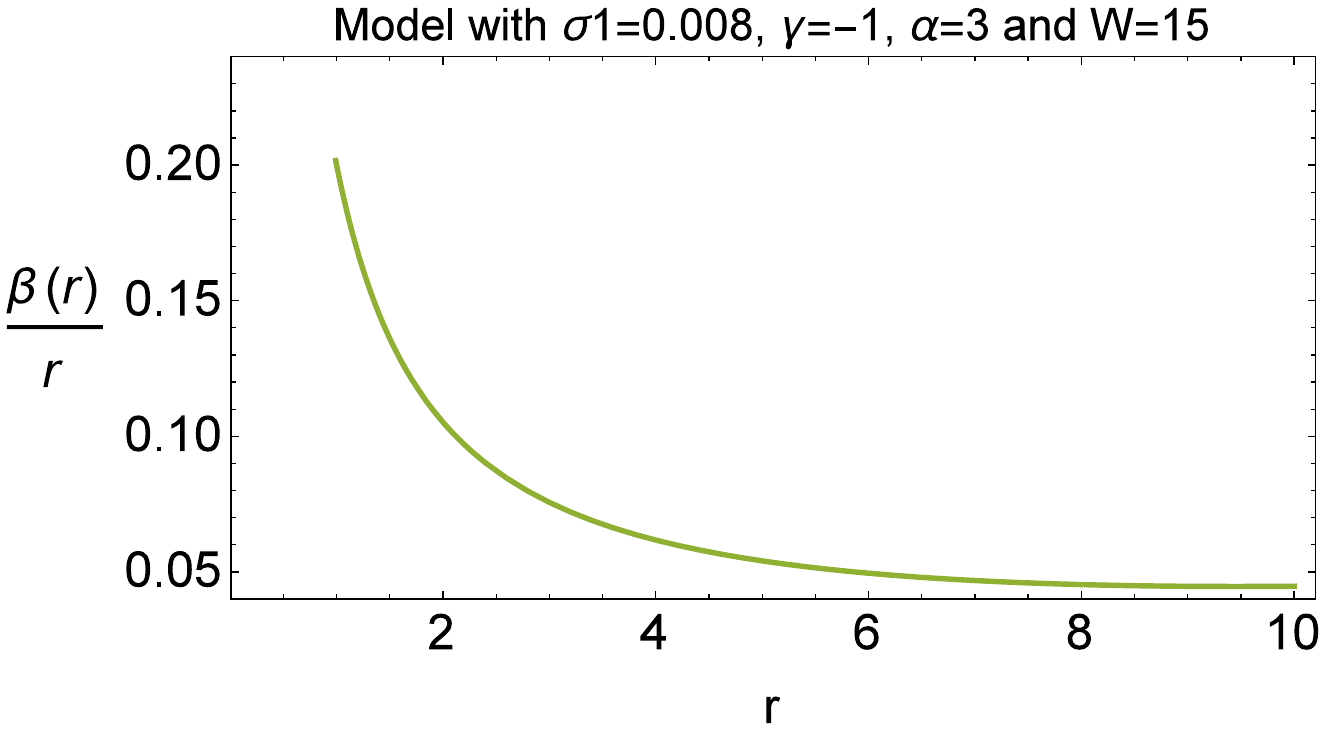}\label{fig11c}}
	\hfill
	\subfloat[Plot of
$\beta(r)-r$]{\includegraphics[width=0.5\textwidth]{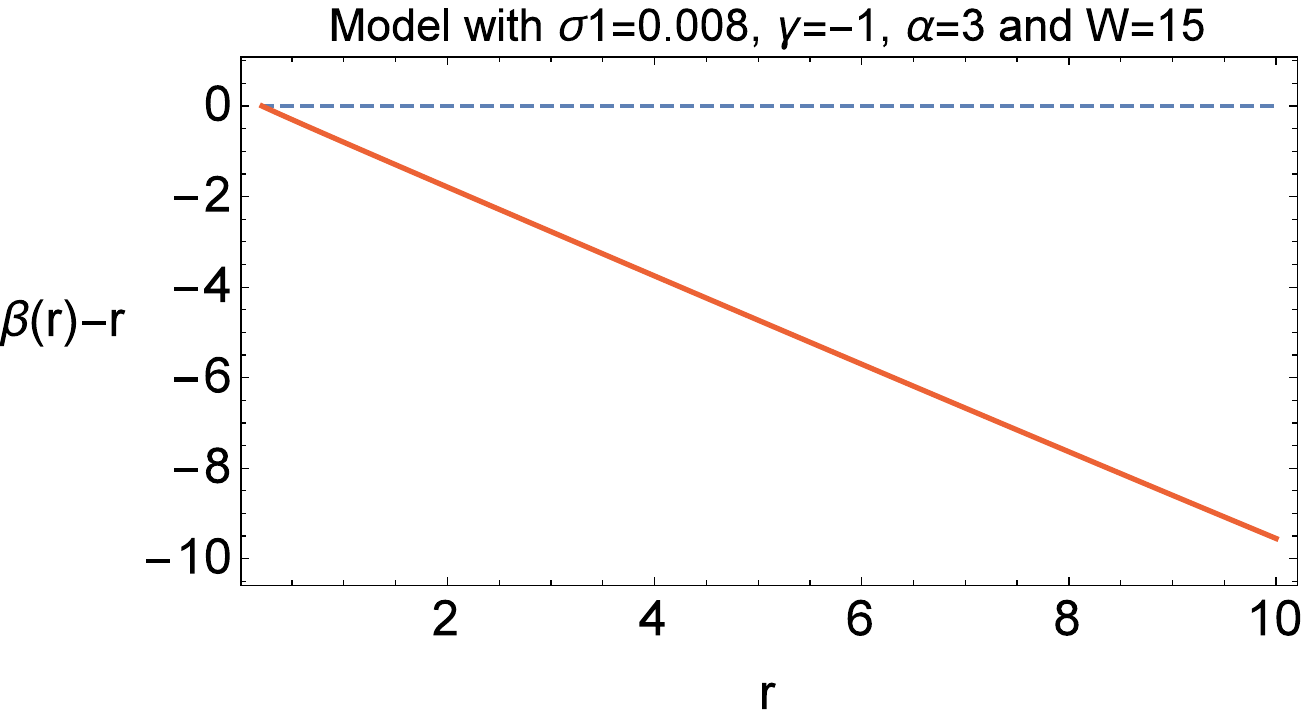}\label{fig11d}}
	\hfill
\caption{The behavior of $\frac{\beta(r)}{r}$ and $\beta(r)-r$ versus $r$ taking
$\omega_0=-2$, $\sigma_1=0.008$, $\gamma=-1$, $\alpha=3$, $W=15$, $\phi_0=10$, $V_0=0.1$.}\label{fig11*}
\end{figure}

\begin{figure}[H]
\centering
{\includegraphics[width=0.7\textwidth]{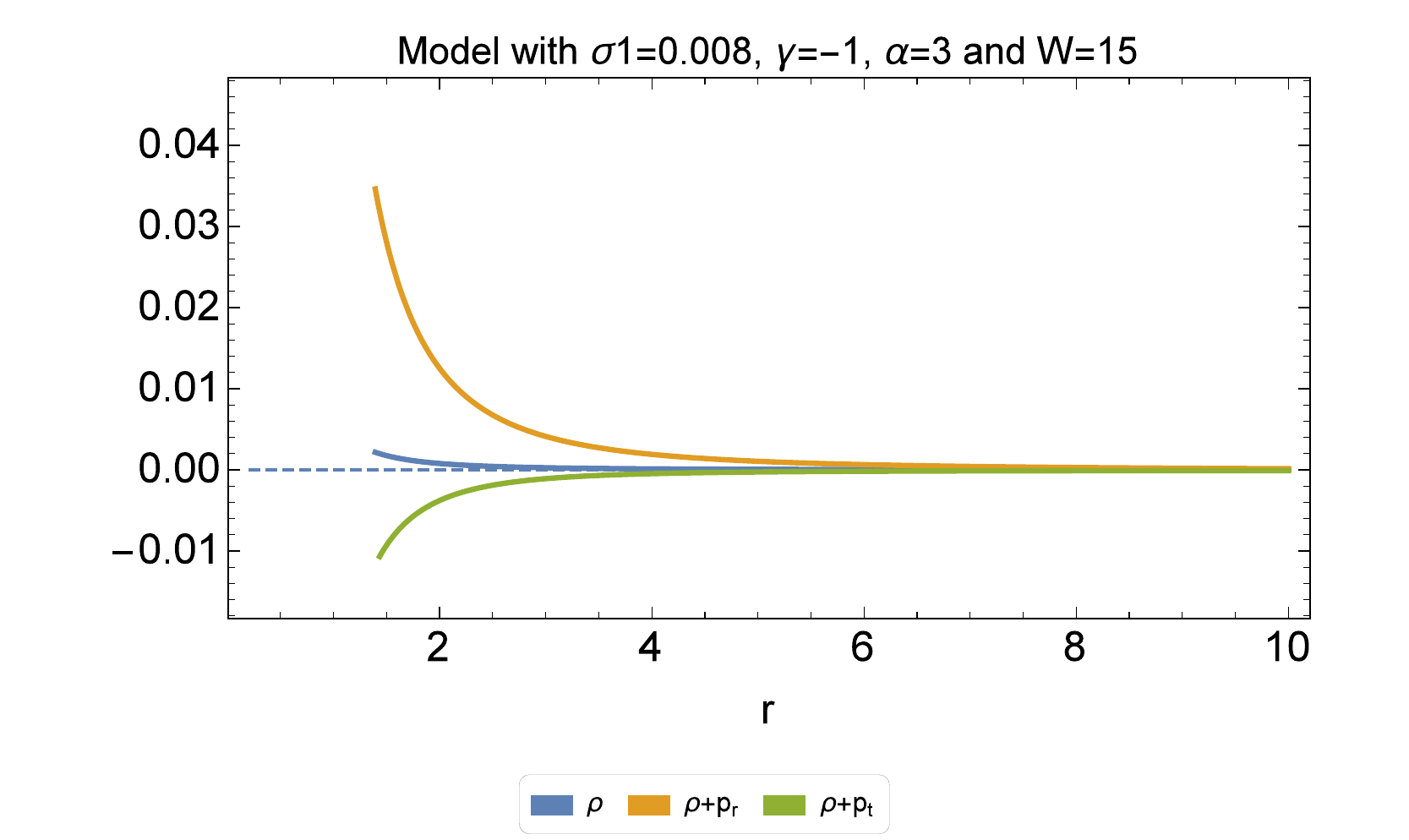}}
\caption{The behavior of $\rho$, $\rho+p_r$ and $\rho+p_t$ versus $r$ for the parameters $\omega_0=-2$, $\sigma_1=0.008$, $\gamma=-1$,
$\alpha=3$, $W=15$, $\phi_0=10$, $V_0=0.1$.}\label{fig12}
\end{figure}

In case of Brans-Dicke we are using $n=1$, $m=-1$ and by varying the parameters $\omega_0$, $\sigma_1$, $\gamma$, $\alpha$, $W$, $\phi_0$, $V_0$ we will discuss the behavior of $\beta(r)$, $\beta'(r)$, $\frac{\beta(r)}{r}$, $\beta(r)-r$, $\rho$, $\rho+p_r$ and $\rho+p_t$. In Fig.~\ref{fig11a}, the behavior of the shape function is shown. This figure shows the increasing behavior and meet the inequality $\beta(r)<r$. It can be seen from Fig.~\ref{fig11c} that $\beta(r)/r
\rightarrow0$ as $r\rightarrow\infty$ which means that the spacetime is asymptotically flat. Fig.~\ref{fig11d} shows that $\beta(r)-r < 0$, which fulfills the condition $1-\beta(r)/r>0$. The throat is then located at $r_0\approx 0.2114$ with $\beta(r_0)=r_0$. The plot of $\beta'(r)$ is shown in Fig.~\ref{fig11b} and
$\beta'(r_0)\approx -0.07093$ which fulfills the condition $\beta'(r_0)<1$. Evolution of the validity of NEC and WEC are shown in Fig.~\ref{fig12}. Here $\rho>0$ and $\rho+p_r>0$ are satisfied throughout the evolution but $\rho+p_t>0$ is not satisfied. Then, the wormhole satisfy NEC-1 and WEC-1 but does not satisfy the full WEC.

\subsubsection{Induced gravity}
\begin{figure}[H]
	\captionsetup{justification=raggedright}
	\subfloat[Plot of
$\beta(r)$]{\includegraphics[width=0.5\textwidth]{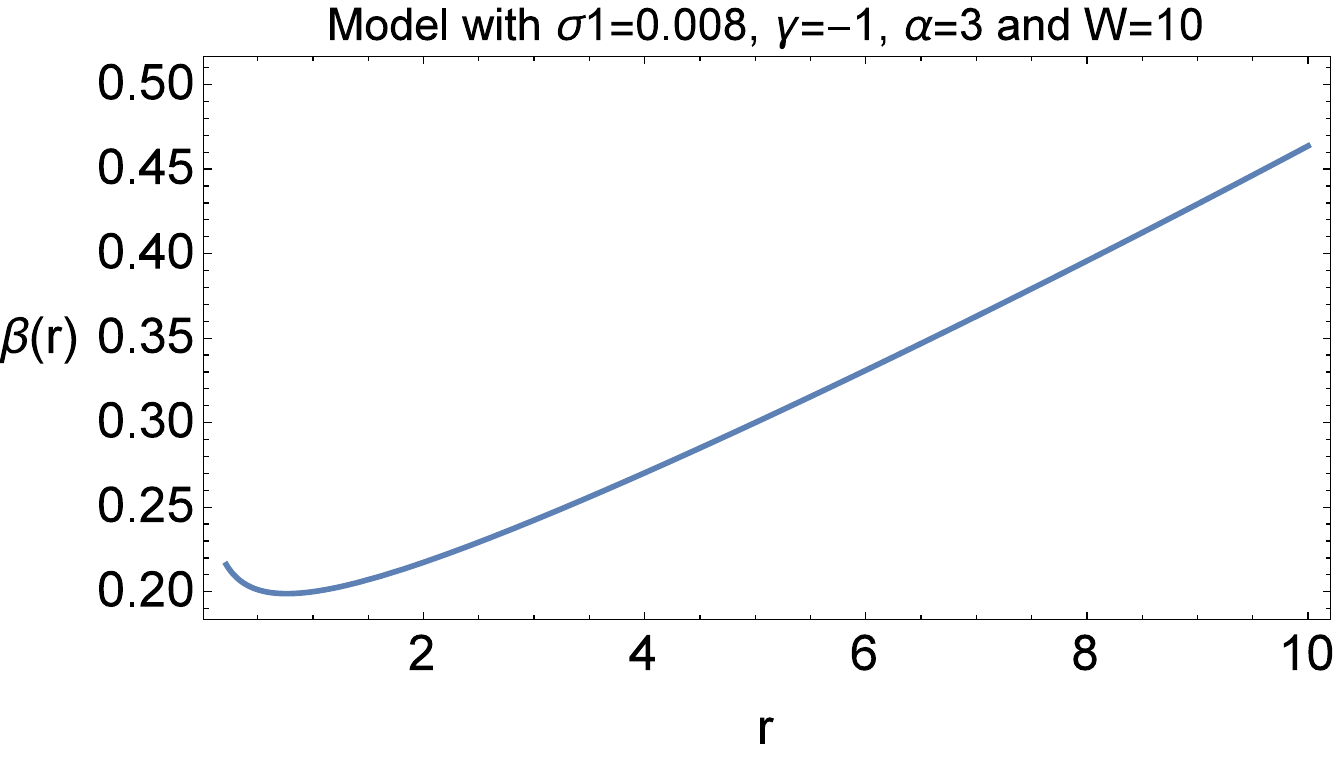}\label{fig13a}}
	\hfill
	\subfloat[Plot of
$\beta'(r)$]{\includegraphics[width=0.5\textwidth]{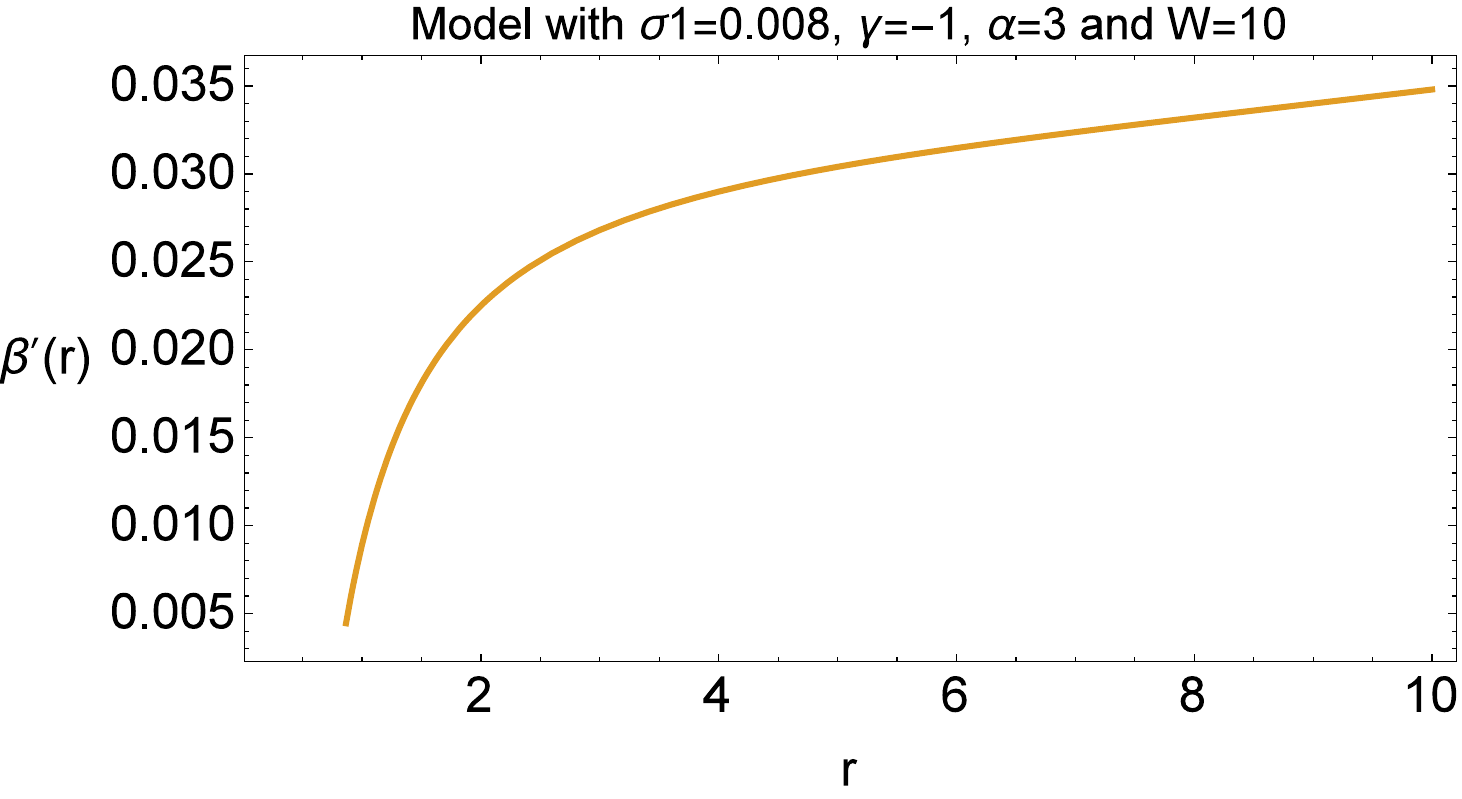}\label{fig13b}}
	\hfill
\caption{The behavior of $\beta(r)$ and $\beta'(r)$ versus $r$ taking
$\omega_0=-2$, $\sigma_1=0.008$, $\gamma=-1$, $\alpha=3$, $W=10$, $\phi_0=10$, $V_0=0.1$.}\label{fig13}
\end{figure}

\begin{figure}[H]
\captionsetup{justification=raggedright}
	\subfloat[Plot of
$\frac{\beta(r)}{r}$]{\includegraphics[width=0.5\textwidth]{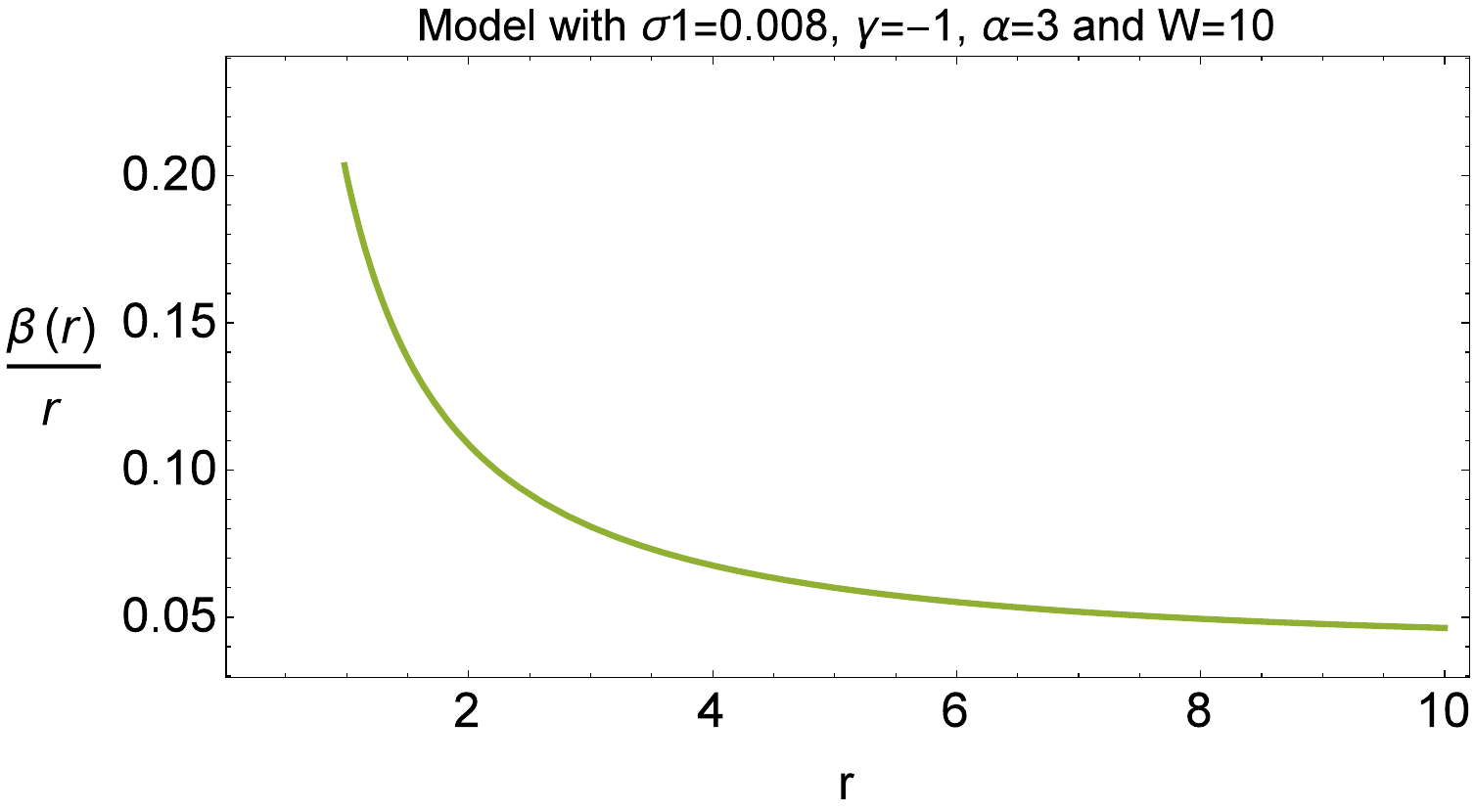}\label{fig13c}}
	\hfill
	\subfloat[Plot of
$\beta(r)-r$]{\includegraphics[width=0.5\textwidth]{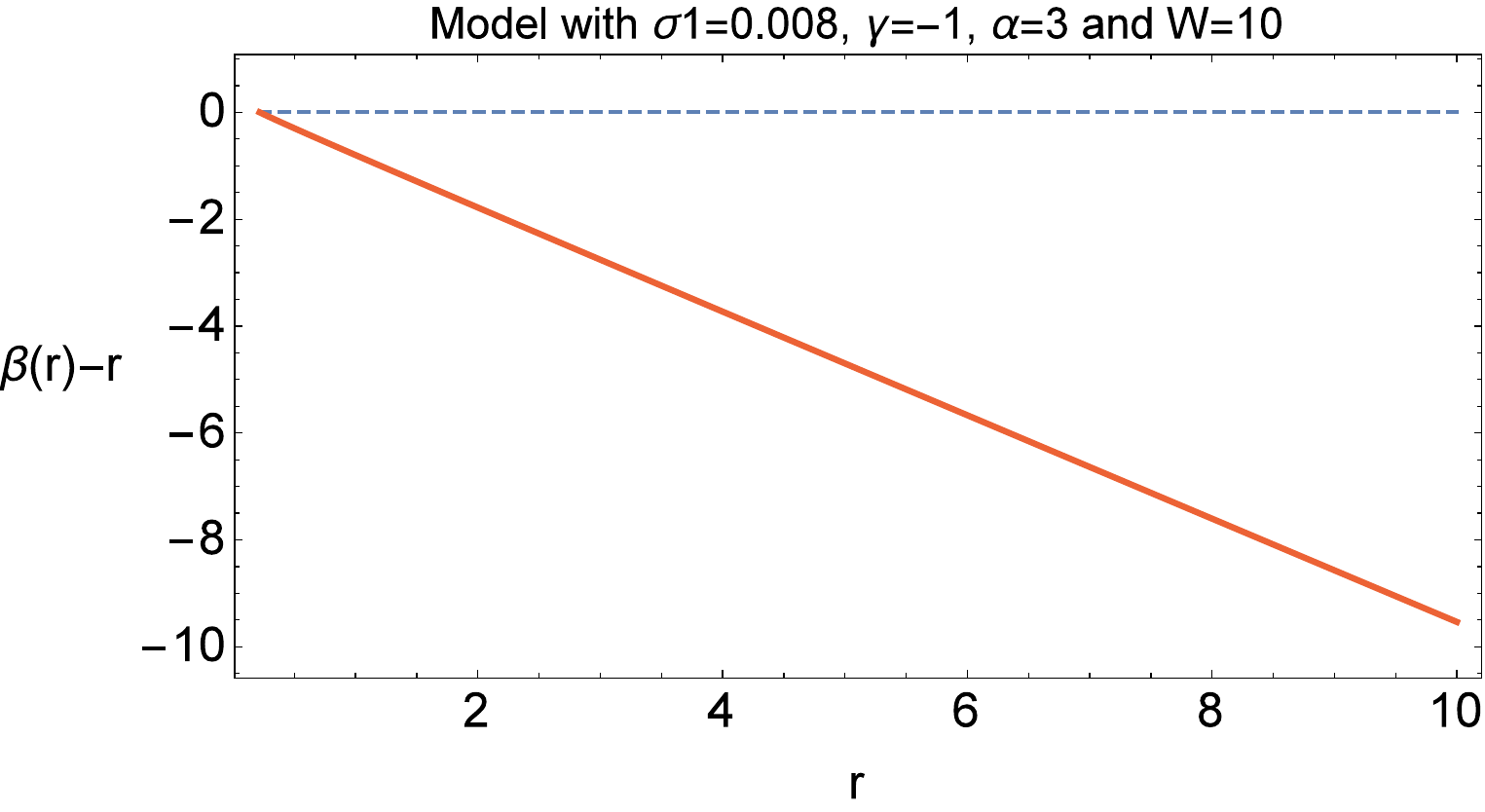}\label{fig13d}}
	\hfill
\caption{The behavior of $\frac{\beta(r)}{r}$ and $\beta(r)-r$ versus $r$ taking
$\omega_0=-2$, $\sigma_1=0.008$, $\gamma=-1$, $\alpha=3$, $W=10$, $\phi_0=10$, $V_0=0.1$.}\label{fig13*}
\end{figure}

\begin{figure}[H]
\centering
{\includegraphics[width=0.6\textwidth]{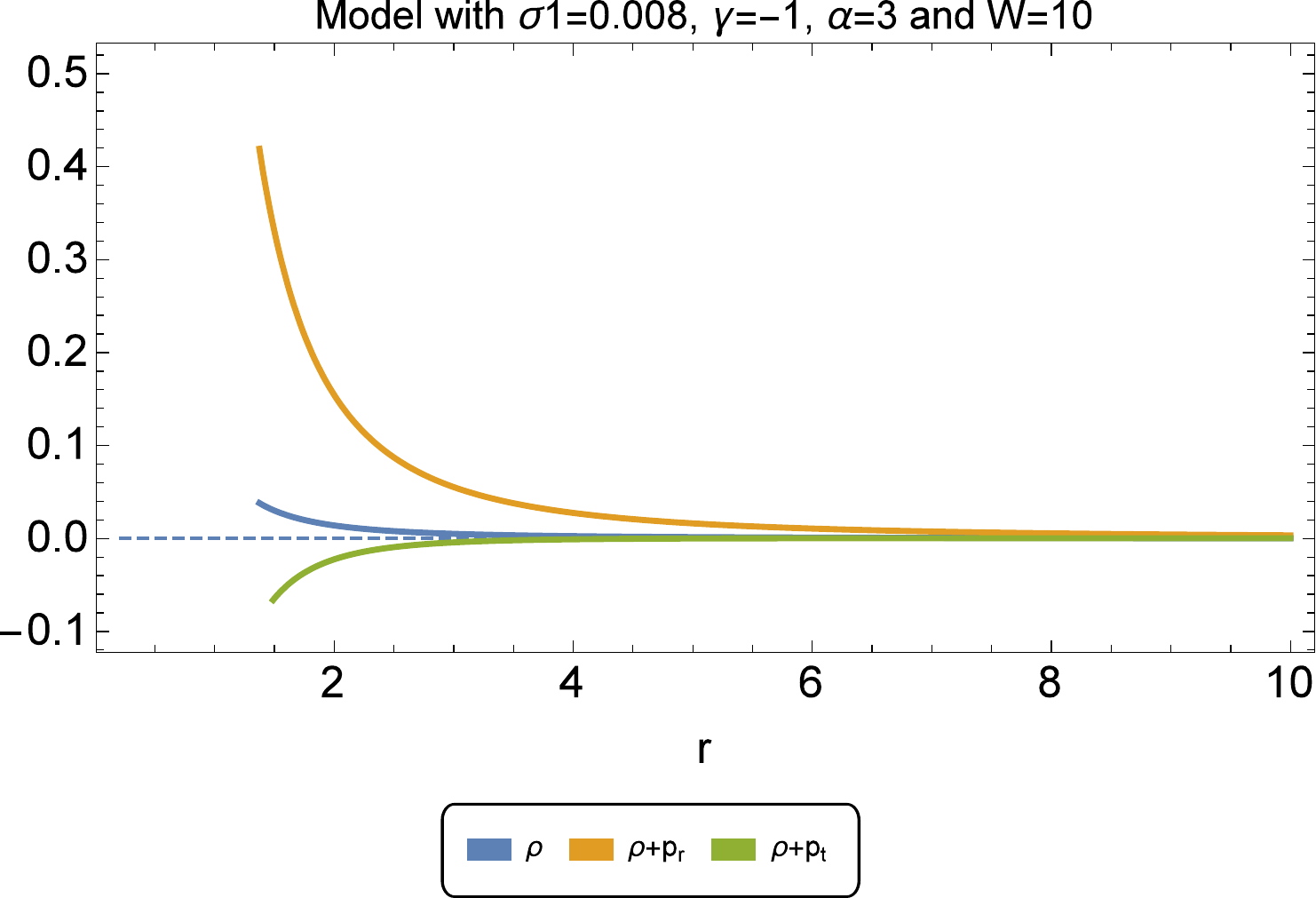}}
\caption{The behavior of $\rho$, $\rho+p_r$ and $\rho+p_t$ for the parameters $\omega_0=-2$,
$\sigma_1=0.008$, $\gamma=-1$, $\alpha=3$, $W=10$, $\phi_0=10$, $V_0=0.1$.}\label{fig14}
\end{figure}

For Induced Gravity, we need to take $n=2$ and $m>0$ and then by varying the parameters $\omega_0$, $\sigma_1$, $\gamma$,
$\alpha$, $W$, $\phi_0$, $V_0$, we will discuss the behavior of $\beta(r)$,
$\beta'(r)$, $\frac{\beta(r)}{r}$, $\beta(r)-r$, $\rho$, $\rho+p_r$ and $\rho+p_t$. In Fig.~\ref{fig13a}, the behavior of the shape
function $\beta(r)$ versus the radial coordinate is plotted. It can be
seen that this function is increasing and then meet the inequality
$\beta(r)<r$. From Fig.~\ref{fig13c}, we can also notice that $\beta(r)/r \rightarrow0$ as
$r\rightarrow\infty$ which means that the spacetime is asymptotically flat. The plot in Fig.~\ref{fig13d} shows that $\beta(r)-r < 0$, which fulfills the condition $1-\beta(r)/r>0$.
The throat is located at $r_0=0.2159$ with $\beta(r_0)=r_0$. The plot of $\beta'(r)$ is shown in Fig.~\ref{fig13b} and $\beta'(r_0)=-0.108932$ which fulfills the
flaring-out condition $\beta'(r_0)<1$. Evolution of NEC and WEC are then shown in Fig.~\ref{fig14}. Here, exactly as in the Brans-Dicke case, $\rho>0$ and $\rho+p_r>0$ are satisfied throughout the
evolution but $\rho+p_t>0$ is not satisfied in this case.

\subsection{ $W(r)=Br^l$}
\begin{figure}[H]
	\captionsetup{justification=raggedright}
	\subfloat[Plot of
$\beta(r)$]{\includegraphics[width=0.5\textwidth]{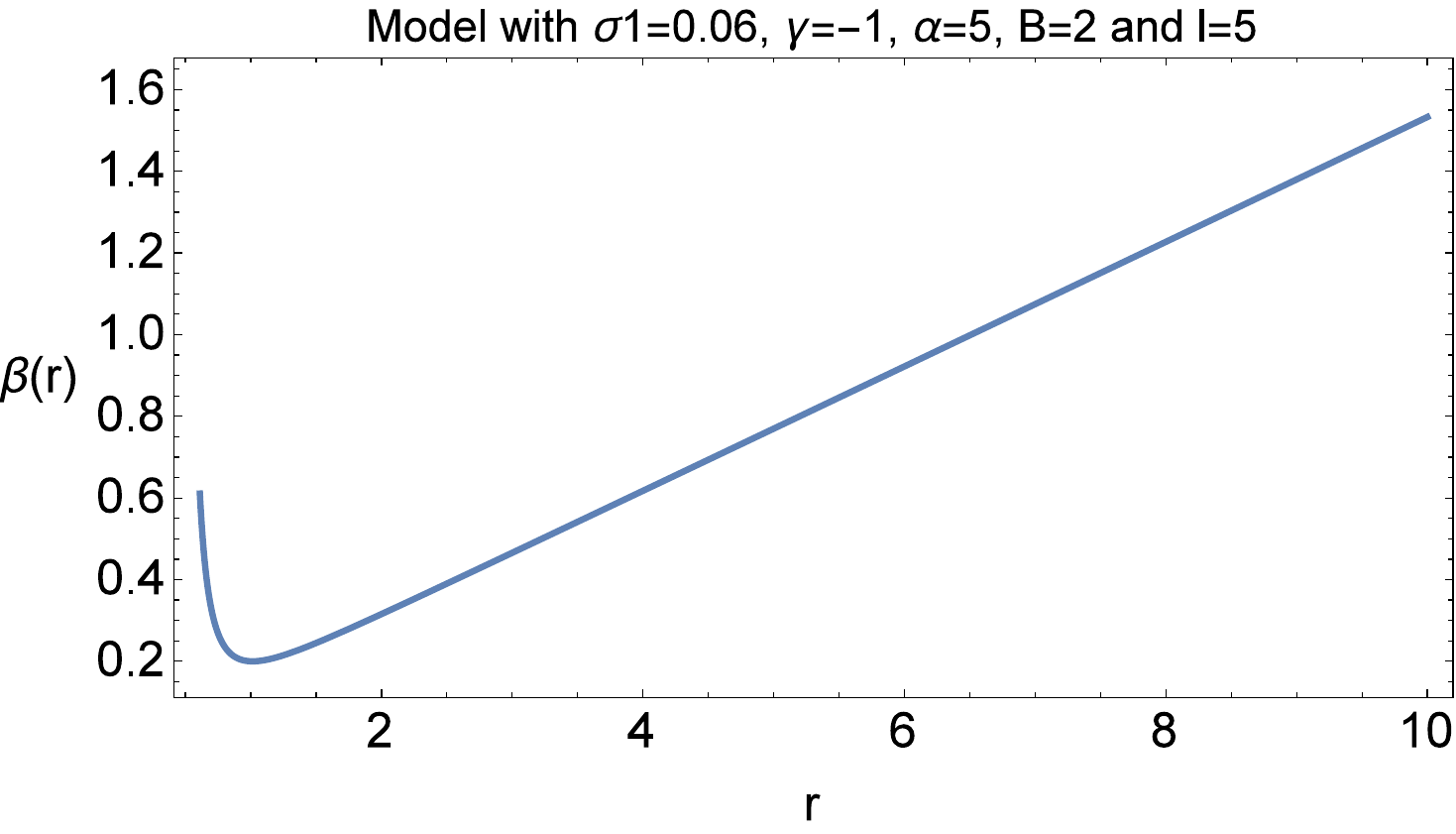}\label{fig15a}}
	\hfill
	\subfloat[Plot of
$\beta'(r)$]{\includegraphics[width=0.5\textwidth]{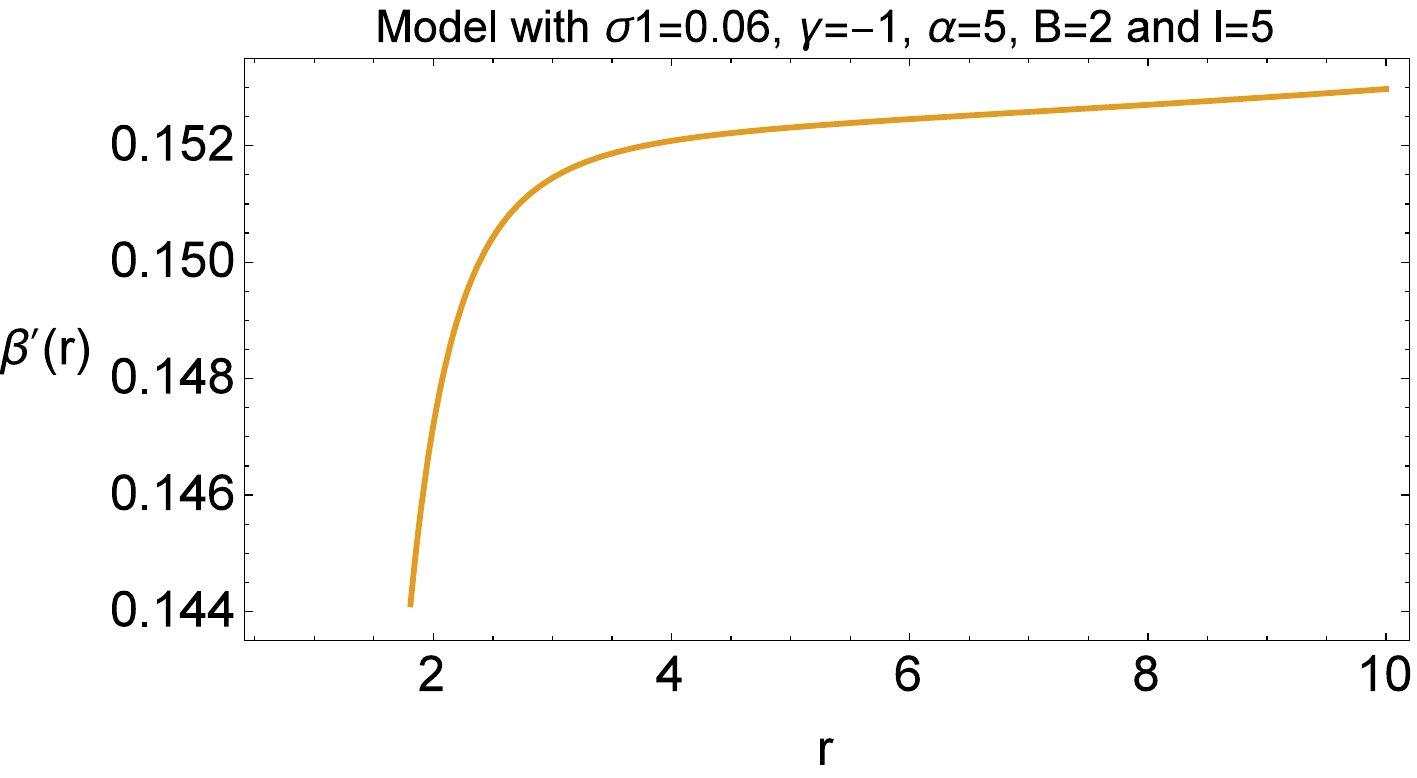}\label{fig15b}}
	\hfill
\caption{Plots of $\beta(r)$ and $\beta'(r)$ versus $r$ for the Brans-Dicke theory taking $\omega_0=-2$,
$\sigma_1=0.06$, $\gamma=-1$, $\alpha=5$, $B=2$, $l=5$ $\phi_0=10$, $V_0=0.1$. In this case we have that the throat is located at $r_0=0.61045$ and then $\beta'(r_0)=-6.11107$. }\label{fig15}
\end{figure}

\begin{figure}[H]
    \captionsetup{justification=raggedright}
	\subfloat[Plot of
$\frac{\beta(r)}{r}$]{\includegraphics[width=0.5\textwidth]{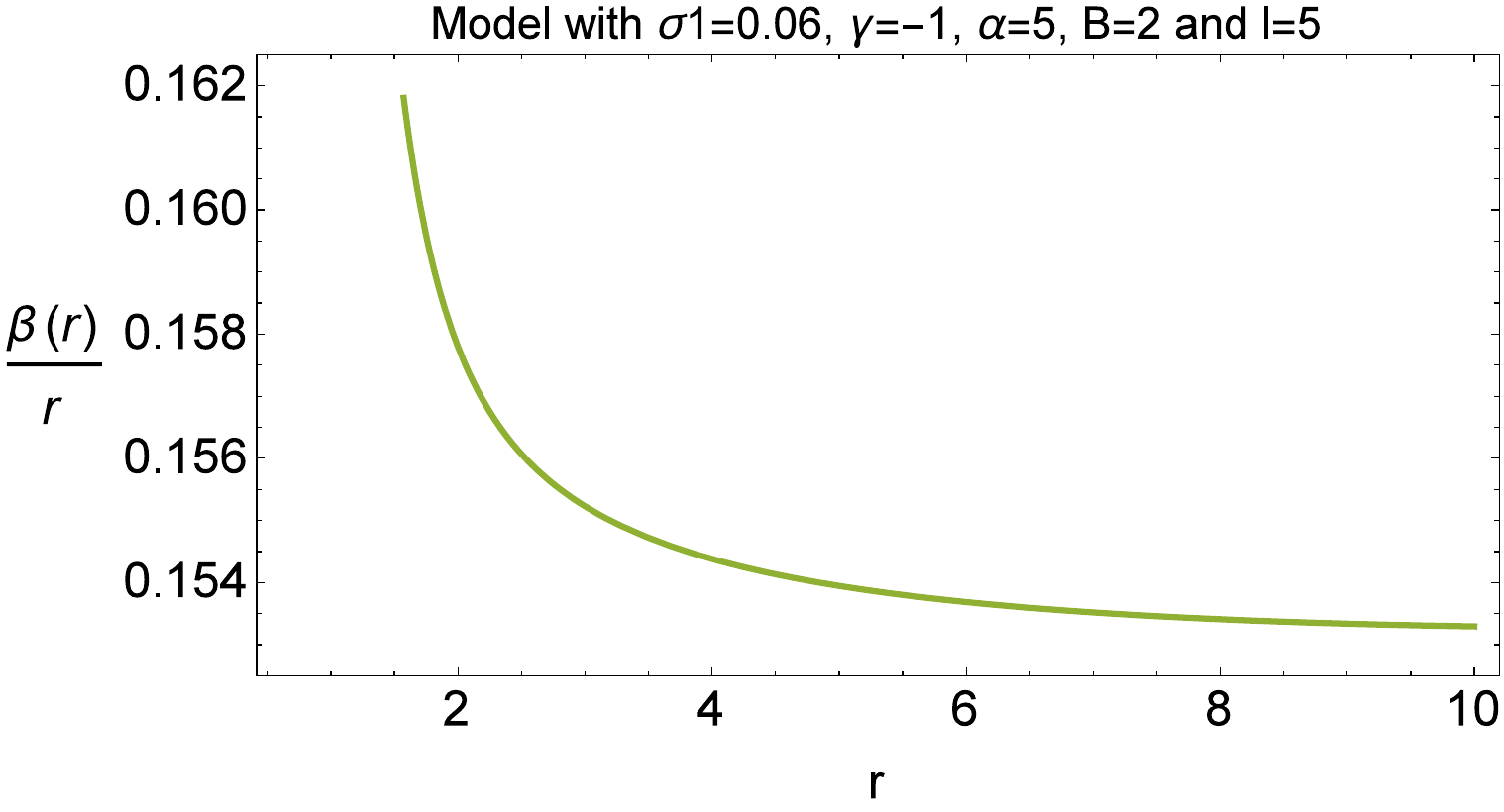}\label{fig15c}}
	\hfill
	\subfloat[Plot of $\beta(r)-r$]{\includegraphics[width=0.5\textwidth]{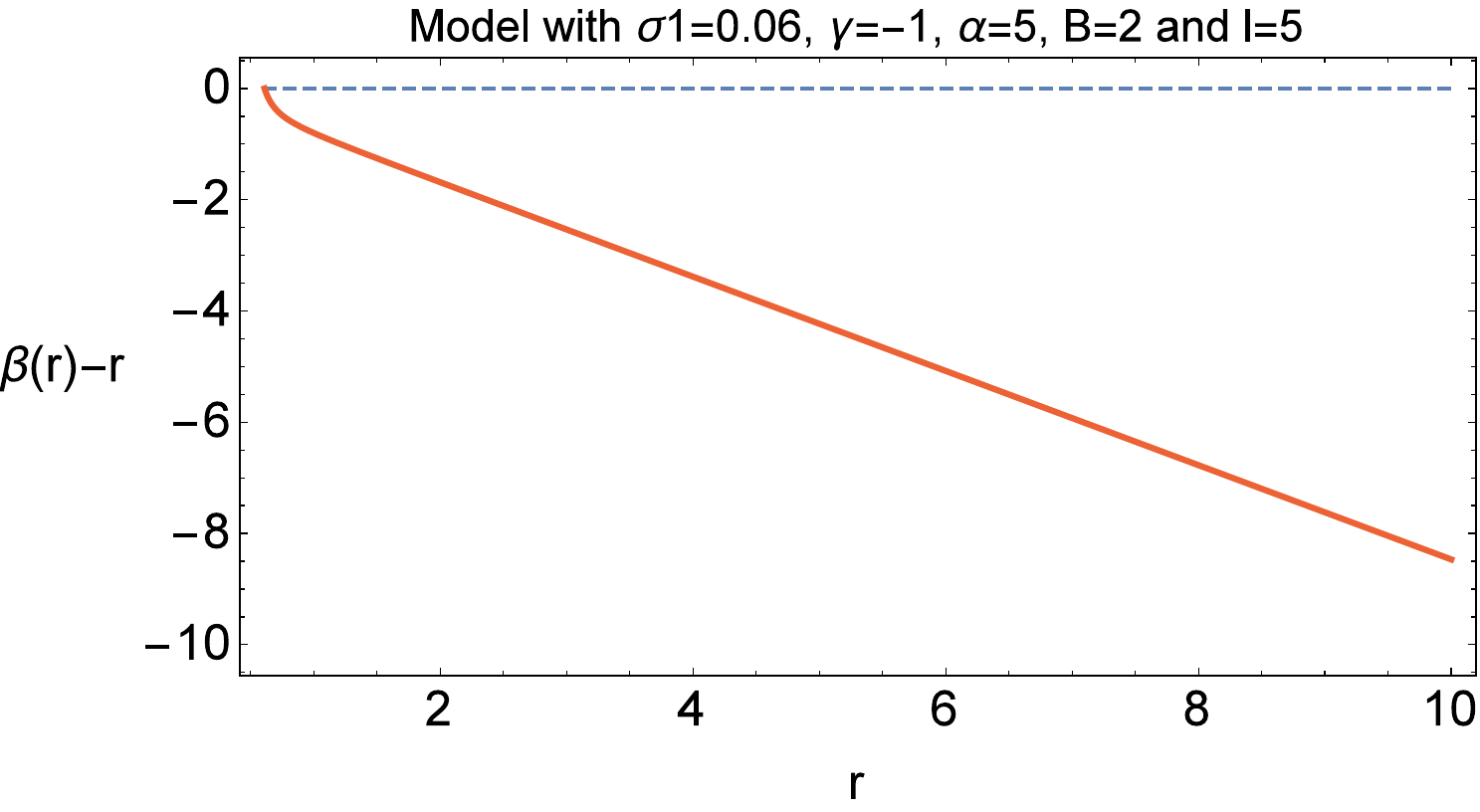}\label{fig15d}}
\caption{Plots of $\frac{\beta(r)}{r}$ and $\beta(r)-r$ versus $r$ for the Brans-Dicke theory for the parameters $\omega_0=-2$,
$\sigma_1=0.06$, $\gamma=-1$, $\alpha=5$, $B=2$, $l=5$ $\phi_0=10$, $V_0=0.1$. In this case the throat is located at $r_0=0.61045$}\label{fig15*}
\end{figure}

\begin{figure}[H]
\centering
{\includegraphics[width=0.6\textwidth]{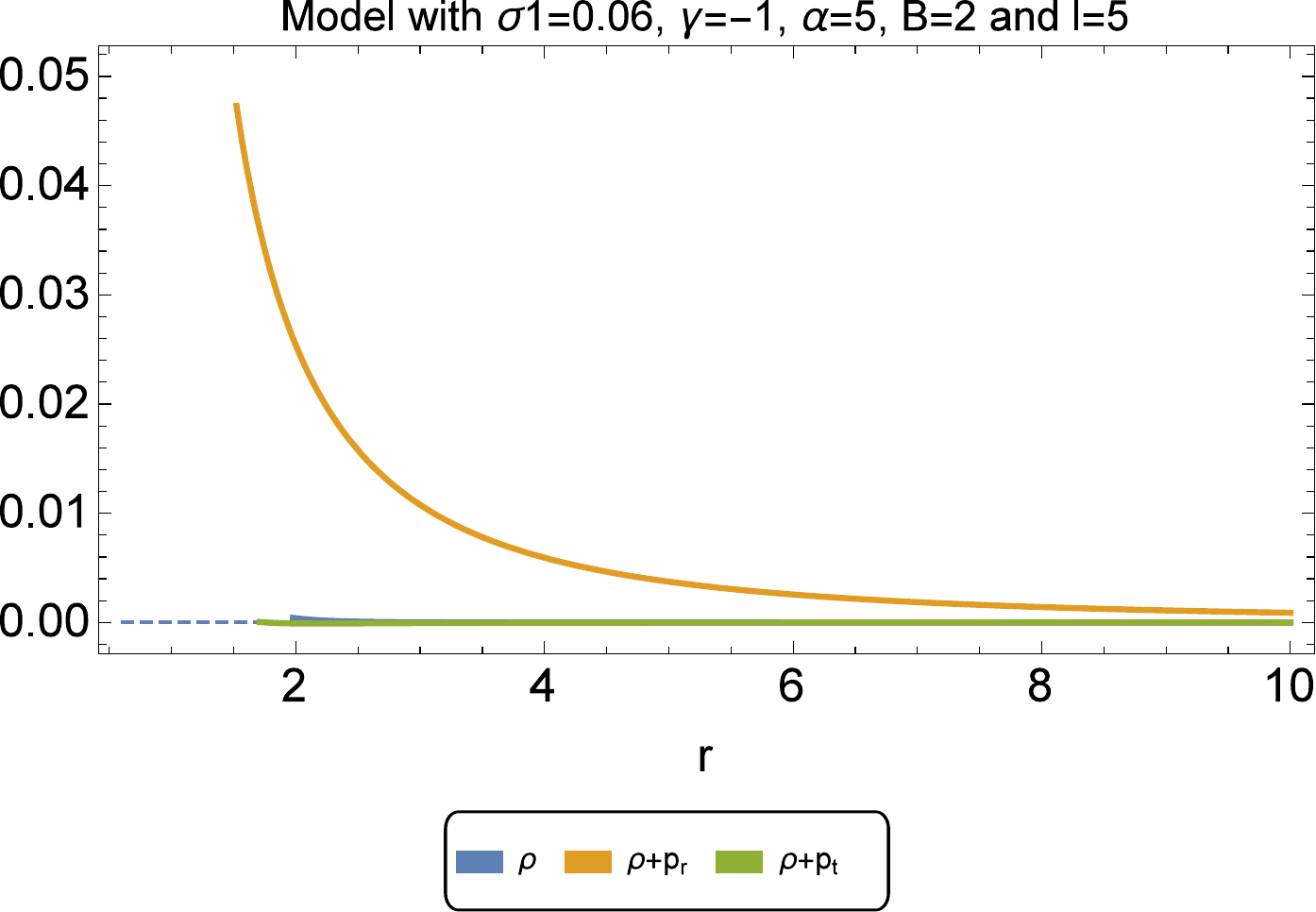}}
\caption{Plots of $\rho$, $\rho+p_r$ and $\rho+p_t$ versus $r$ for the Brans-Dicke theory for the parameters $\omega_0=-2$, $\sigma_1=0.008$, $\gamma=-1$, $\alpha=5$, $B=2$, $l=5$ $\phi_0=10$, $V_0=0.1$. In this case the throat is located at $r_0=0.61045$}\label{fig16}
\end{figure}

\begin{figure}[H]
	\captionsetup{justification=raggedright}
	\subfloat[Plot of
$\beta(r)$]{\includegraphics[width=0.5\textwidth]{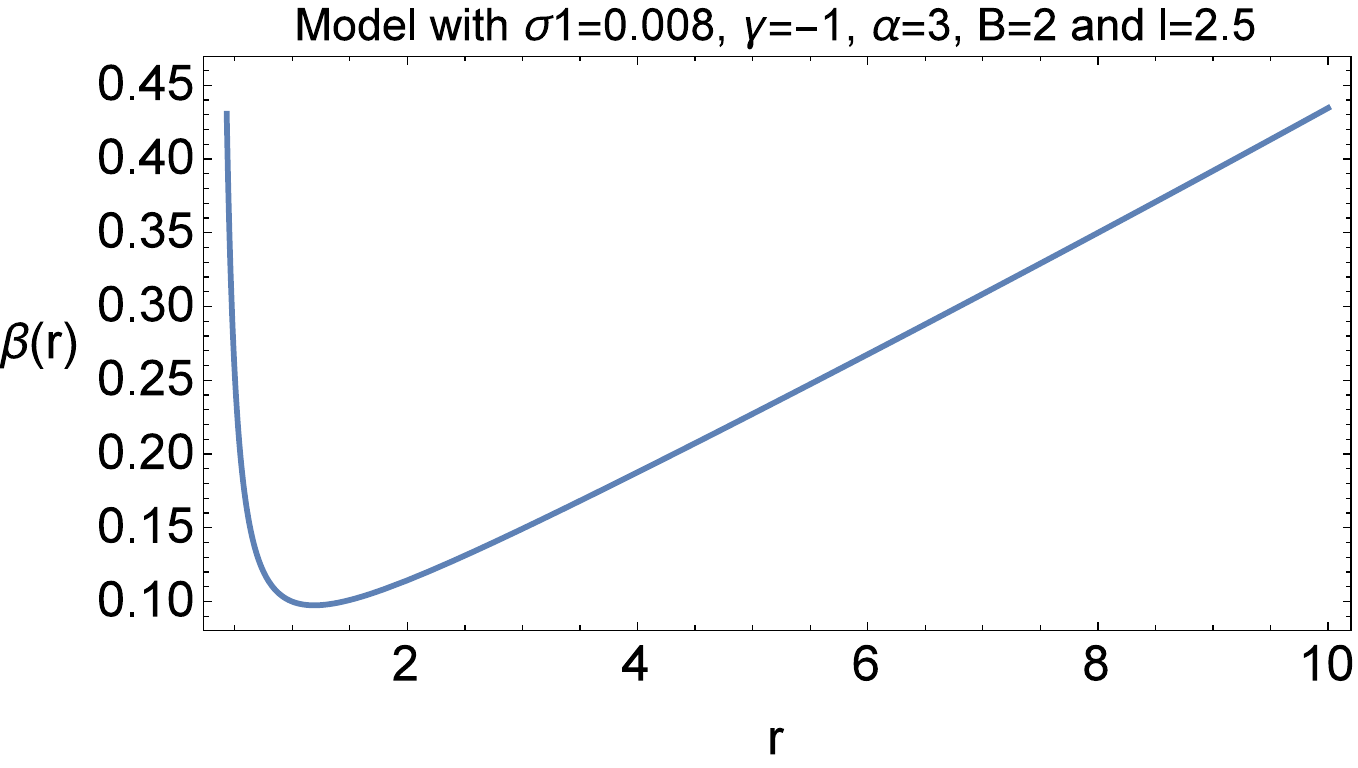}\label{fig17a}}
	\hfill
	\subfloat[Plot of
$\beta'(r)$]{\includegraphics[width=0.5\textwidth]{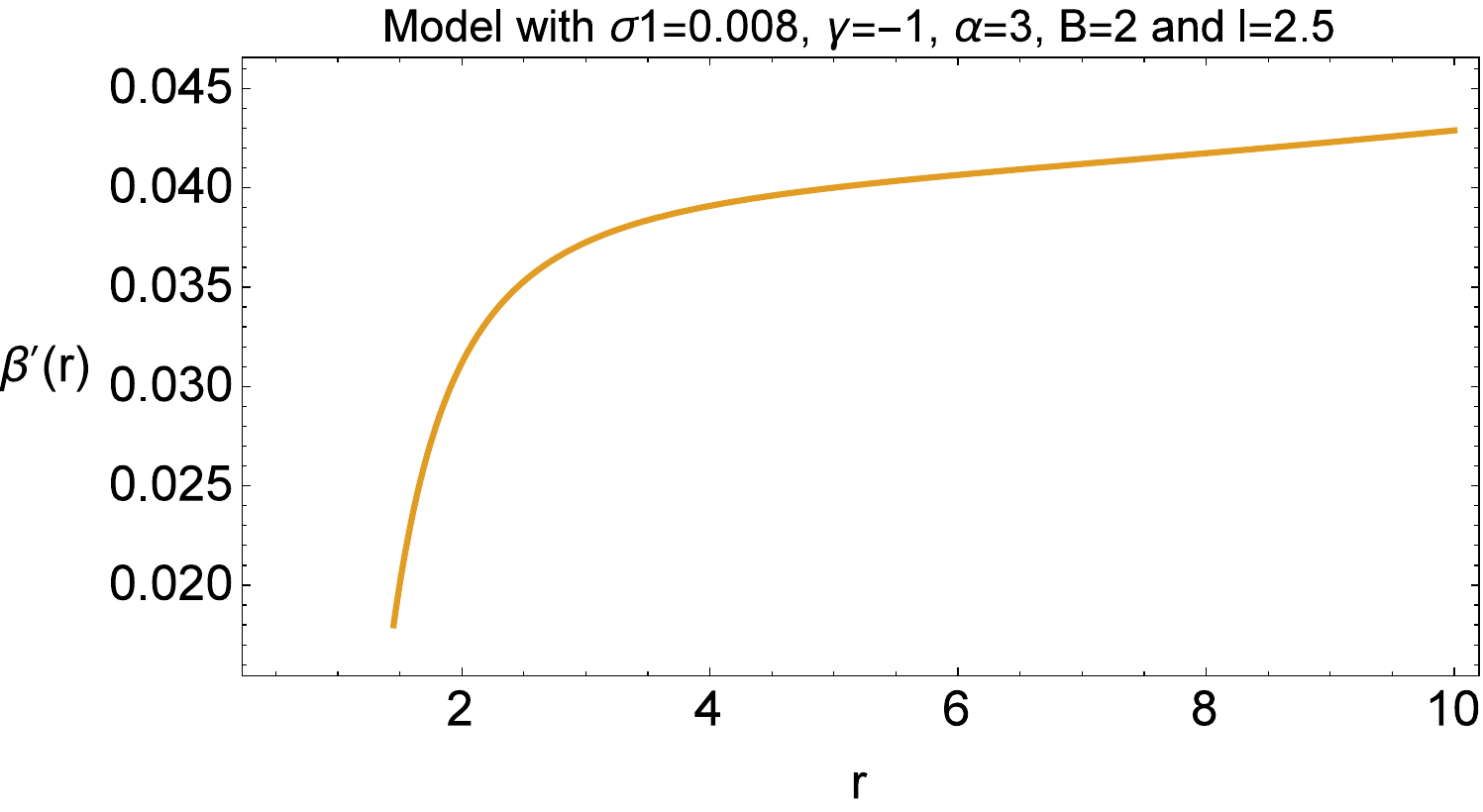}\label{fig17b}}
	\hfill
\caption{Plots of $\beta(r)$, $\beta'(r)$ and $\frac{\beta(r)}{r}$ versus $r$ for Induced Gravity taking
$\omega_0=-2$, $\sigma_1=0.008$, $\gamma=-1$, $\alpha=3$, $B=2$, $l=2.5$ $\phi_0=10$,
$V_0=0.1$. Here we have that the throat is located at $r_0=0.43055$ and also $\beta'(r_0)=-4.15156$ }\label{fig17}
\end{figure}

\begin{figure}[H]
	\captionsetup{justification=raggedright}
	\subfloat[Plot of
$\frac{\beta(r)}{r}$]{\includegraphics[width=0.5\textwidth]{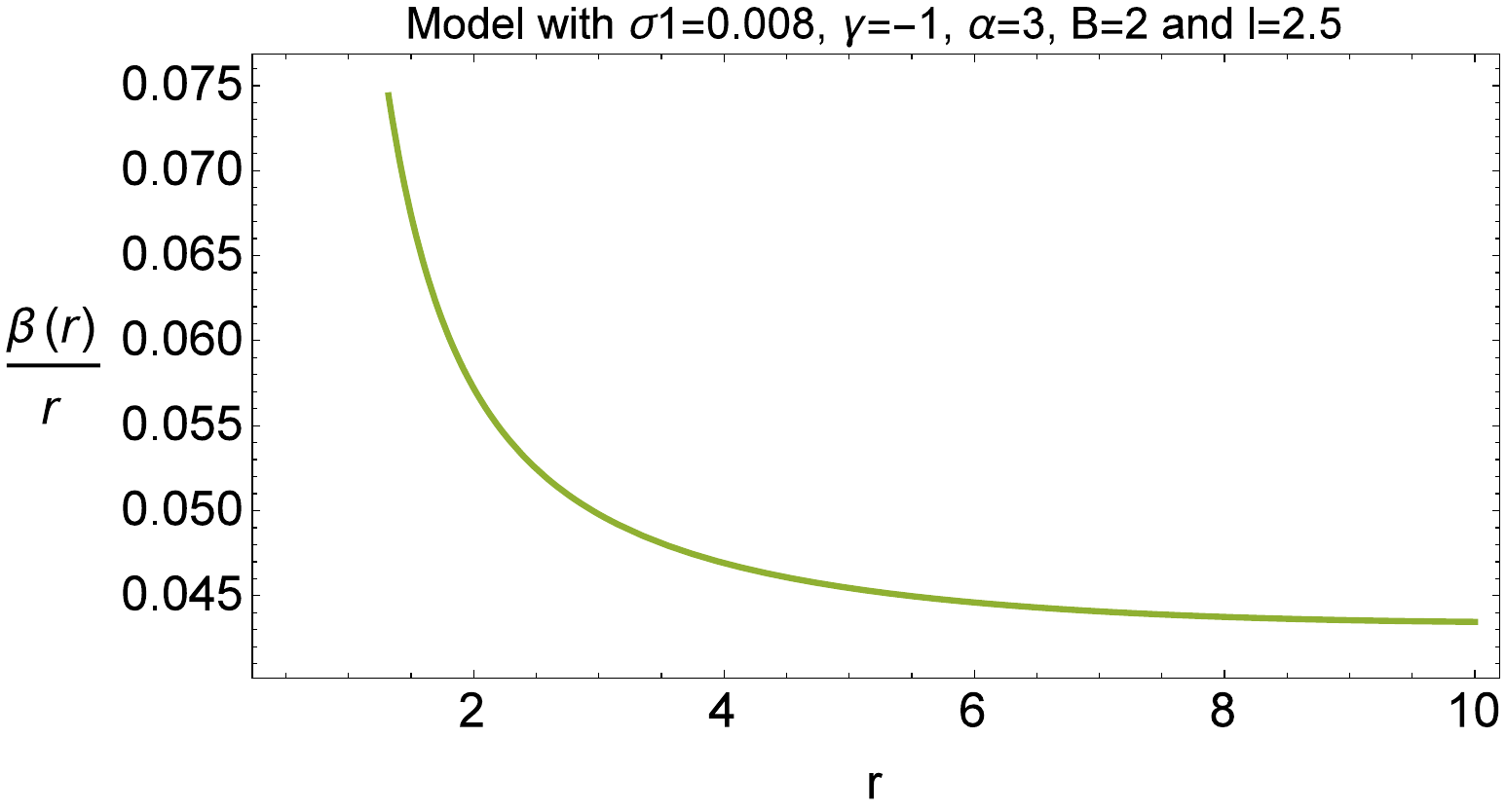}\label{fig17c}}
	\hfill
	\subfloat[Plot of
$\beta(r)-r$]{\includegraphics[width=0.5\textwidth]{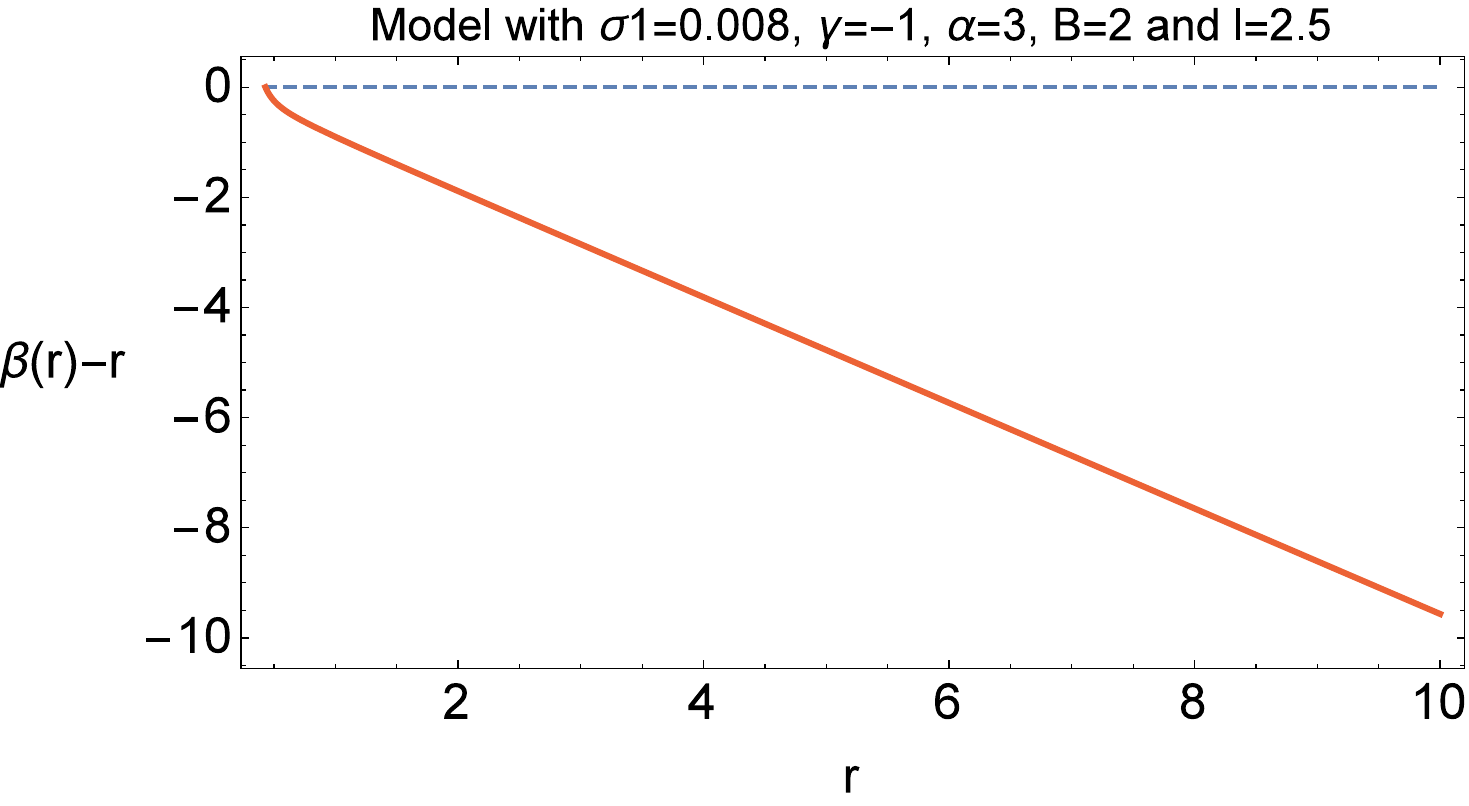}\label{fig17d}}
	\hfill
\caption{Plots of $\beta(r)$, $\beta'(r)$ and $\frac{\beta(r)}{r}$ versus $r$ for induced gravity  taking
$\omega_0=-2$, $\sigma_1=0.008$, $\gamma=-1$, $\alpha=3$, $B=2$, $l=2.5$ $\phi_0=10$,
$V_0=0.1$. Here we have that the throat is at $r_0=0.43055$}\label{fig17*}
\end{figure}

\begin{figure}[H]
\centering
{\includegraphics[width=0.7\textwidth]{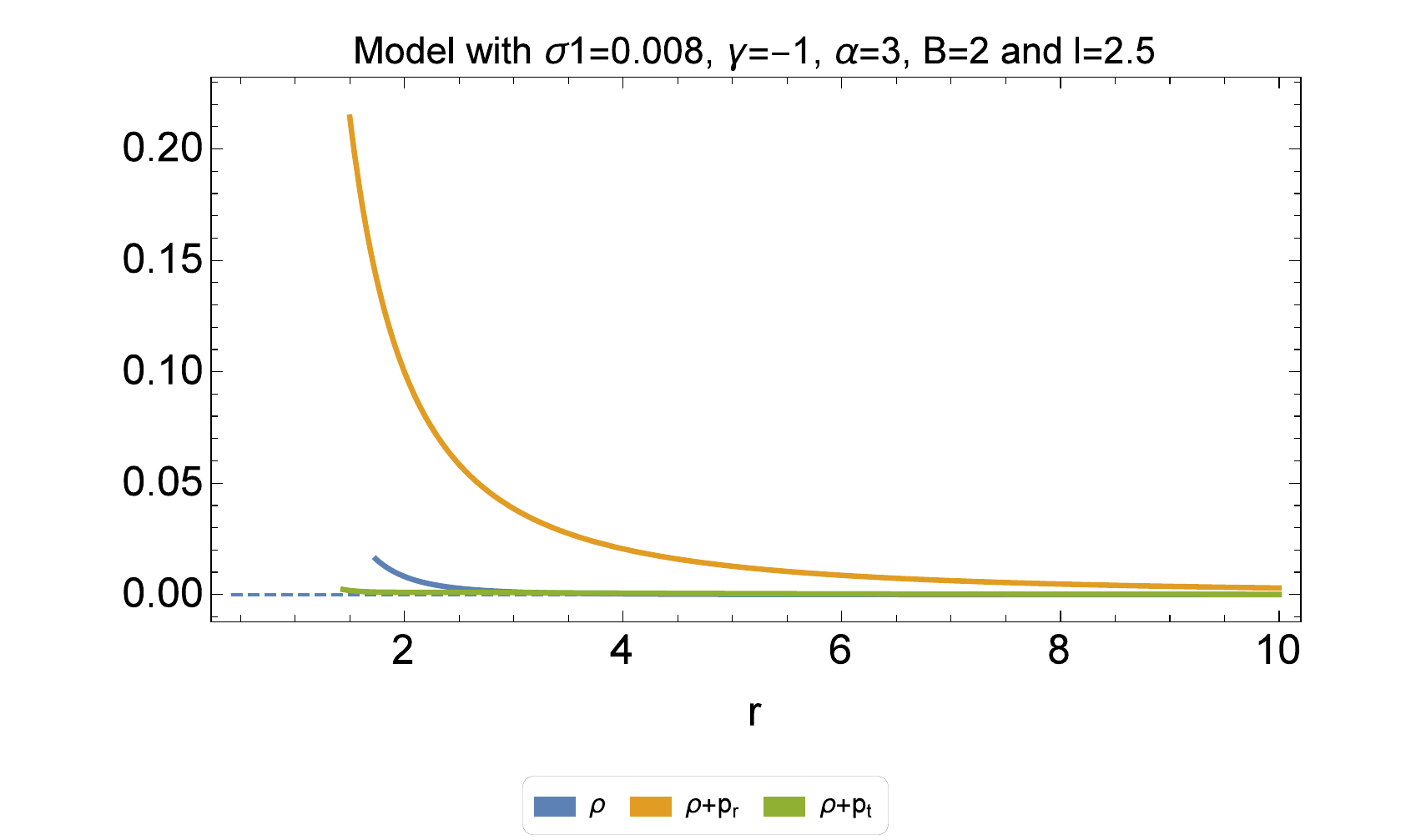}}
\caption{Plots show the evolution of
$\rho$, $\rho+p_r$ and $\rho+p_t$ for for Induced Gravity for the parameters $\omega_0=-2$,
$\sigma_1=0.008$, $\gamma=-1$, $\alpha=3$, $B=2$, $l=2.5$ $\phi_0=10$, $V_0=0.1$. Here we have that the throat is at $r_0=0.43055$}\label{fig18}
\end{figure}

Let us now explore the case where we take $W(r)=Br^l$ with $B$ and $l$ being positive constants. By using this EoS in the field equations (\ref{2.12})-(\ref{2.14}), we get a constraint of the
following form
\begin{eqnarray}
\nonumber&&\frac{1}{r\kappa^2}\left(\frac{d}{r}\right)^{-\alpha\sigma_1}{\phi_0}^{-\alpha}
\bigg[-2\kappa^2r^3V_0(1+Br^l)+2nr\gamma\sigma_1(-2+3Br^l+n\sigma_1Br^l){\phi_0}^{n+\alpha}
\left(\frac{d}{r}\right)^{\sigma_1(n+\alpha)}\\
\nonumber&&-r\omega_0{\sigma_1}^2(1+Br^l){\phi_0}^{m+2+\alpha}\left(\frac{d}{r}\right)^{\sigma_1
(m+2+\alpha)}+\omega_0{\sigma_1}^2(1+Br^l){\phi_0}^{m+2+\alpha}\left(\frac{d}{r}\right)^{\sigma_1
(m+2+\alpha)}\beta(r)\\
\nonumber&&-\gamma{\phi_0}^{n+\alpha}\left(\frac{d}{r}\right)^{\sigma_1(n+\alpha)}
\left\{(2-4n\sigma_1+7n\sigma_1Br^l+2n^2{\sigma_1}^2Br^l)\beta(r)+rBr^l(2-n\sigma_1)\beta'(r)
\right\}\bigg]=0\,.
\end{eqnarray}
Since this equation is very complicated to solve analytically, we will again solve this equation numerically for Brans-Dicke and Induced gravity cases. For Brans-Dicke theory we will choose the parameters $\omega_0=-2$,
$\sigma_1=0.06$, $\gamma=-1$, $\alpha=5$, $B=2$, $l=5$ $\phi_0=10$, $V_0=0.1$. On the other hand, for Induced Gravity we set $n=2$, $m>0$, $\omega_0=-2$,
$\sigma_1=0.008$, $\gamma=-1$, $\alpha=3$, $B=2$, $l=2.5$ $\phi_0=10$ and $V_0=0.1$. In Figs.~\ref{fig15a} and \ref{fig17a} are depicted the shape function for the Brans-Dicke theory and Induced Gravity respectively. We can see that the shape functions satisfy the required condition
$\beta(r)<r$ for both cases. The graphs in Figs.~\ref{fig15c} and ~\ref{fig17c} show that $\beta(r)/r
\rightarrow0$ as $r\rightarrow\infty$, so that, again the metric is asymptotically flat in Brans-Dicke and Induced Gravity. Figs.~\ref{fig15d} and \ref{fig17d} show that
$\beta(r)-r < 0$ for both theories, which fulfills the condition $1-\beta(r)/r>0$. The plot of $\beta'(r)$ is shown in Fig.~\ref{fig15b} and \ref{fig17b} for both theories where it can be notice that the flaring-out condition is also satisfied for both cases. Finally, Fig.~\ref{fig16} and \ref{fig18} show the evolution of the validity of NEC and WEC for both theories. In these two theories we again have the same situation that $\rho>0$ and $\rho+p_r>0$ are satisfied throughout the evolution but $\rho+p_t>0$ is not satisfied.


\section{Summary and Conclusion}\label{sec4}
Wormhole solutions in GR do not satisfy all the standard energy
conditions. Other approach is then to modify Einstein field equations in terms of an effective energy-momentum tensor that satisfy the energy bounds and
the exotic part of the wormholes are supported by higher order curvature terms.

In this paper, we studied whether in $f(R,\phi)$ modified
theory, the ordinary matter can support wormholes. In the last decades, it has been mentioned that in highly
compacted astrophysical objects,  pressures are anisotropic, which means that the
tangential and radial pressures are not equal for such objects. Investigating the existence of wormholes for different kind of fluids are then an interesting question to address. To investigate this we have
analyzed the behavior of NEC and WEC for three different supporting fluids: a barotropic fluid, an anisotropic fluid and an isotropic fluid. Additionally, we have constructed wormholes satisfying the flaring-out condition $\beta'(r=r_0)<1$ where $r_0$ is the throat which satisfies $\beta(r=r_0)=r_0$.

To find analyse the physics of wormhole solutions, different methods have been discussed in the literature. One method is to
find wormhole solutions giving a specific shape function. On the other hand, a second approach is considering the matter content and then calculate the shape function directly from the field equations.

In our manuscript, we have explored $f(R,\phi)$ gravity involving coupling
between the Ricci scalar and matter field. Here the resulting equations are
highly non-linear and complicated involving unknowns $p_t$, $p_r$, $\rho$,
$a$, $b$, $f(R,\phi)$. Therefore, we have focused our study in a power-law case $f(R,\phi)=\gamma R
\phi^{n}$, where $\gamma$ and $n$ are constants. Then, by choosing $n=1$ and $m=-1$, Brans-Dicke theory is recovered and
by choosing $n=2$ and $m>0$, Induced Gravity is recovered. Then, for an anisotropic, isotropic and barotropic fluids, we have constructed wormhole solutions and then explore the energy conditions.

In the case of an anisotropic matter content, we assumed a specific form of the shape function (power-law type) to obtain a solution and then to check the existence of wormholes. Then, we investigated the validity of standard energy conditions. We have found that in both Brans-Dicke and induced gravity theories, an anisotropic generic fluid verifying all the energy conditions can support a wormhole geometry. However, to satisfy all the energy conditions, we must have negative values of $\gamma$. For positive values of $\gamma$, the matter given by the anisotropic fluid will violate some of the energy conditions.

In the case of an isotropic fluid $p_r=p_t=p$, it is possible to solve the field equations to get the shape function and the potential. The shape function then can be constraint to satisfy the wormhole's conditions. This is valid for a generic power-law $f(R,\phi)$ gravity. Then, we found that isotropic fluids satisfying all the energy conditions (WEC and NEC) can support wormholes in both Brans-Dicke and induced gravity theories.

For an anisotropic matter satisfying a barotropic EoS $p_r=W(r)\rho$, the field equations are complicated to solve. Therefore, we studied some special cases numerically focusing on Brans-Dicke and induced gravity theories. This study was carried out by choosing some special values of the parameters. We analised the cases where the barotropic function is a constant $W(r)=W$ and also when $W(r)=Br^l$, where $B$ and $l$ are constants. For both cases, we have constraint the parameters in such a way that ensures the conditions to have a traversable wormhole geometry. In both cases, we have found some models where $\rho+p_r>0$ and also $\rho>0$ but $\rho+p_t$ can be negative, so that WEC is not always satisfy. Then, for our potential, barotropic fluids in Brans-Dicke and Induced Gravity do not satisfy all the energy conditions to support a wormhole geometry. Note that one can also assume a barotropic EoS $p_t=W(r)\rho$, when now the transverse pressure is related to the energy density. If ones carries out the same analysis mentioned above with the same potential, we also get a similar conclusion: wormholes can be constructed satisfying all the geometric required properties but the full WEC is not satisfy. For this case, one has that $\rho+p_t>0$ and also $\rho>0$ but $\rho+p_r$ can be negative.

\vspace{.25cm}

\begin{acknowledgments}
	S.B. is supported by the Comisi{\'o}n Nacional de Investigaci{\'o}n Cient{\'{\i}}fica y
Tecnol{\'o}gica (Becas Chile Grant No.~72150066).
\end{acknowledgments}

%
%
%
%

\end{document}